\documentclass[a4paper,useAMS,usenatbib]{mn2e} 

\usepackage{psfig}
\usepackage{graphicx}
\usepackage{aas_macros}
\usepackage[totalwidth=480pt, totalheight=680pt]{geometry}
\usepackage{amssymb}   
\usepackage{setspace}  

\newlength{\plotwidth}
\newlength{\fullwidth}
\newcommand{\sqdeg}{\mbox{\,deg$^{2}$}}

\newcommand{\kiloparsec}{\mbox{$\,h^{-1}\,{\rm kpc}$}}
\newcommand{\cubicMpc}{\mbox{$\,h^{-3}\,{\rm Mpc}^3$}}

\newcommand{\invcubicMpc}{\mbox{$\,h^3\,{\rm Mpc}^{-3}$}}

\newcommand{\kms}{\mbox{\,km\,s$^{-1}$}}
\newcommand{\kmsMpc}{\mbox{\,km\,s$^{-1}$\,Mpc$^{-1}$}}

\newcommand{\kb}{\mbox{$K$}}

\newcommand{\bj}{\mbox{$b_{\rm\scriptscriptstyle J}$}}
\newcommand{\rf}{\mbox{$r_{\rm\scriptscriptstyle F}$}}

\newcommand{\khjrb}{\mbox{$KHJr_{\rm\scriptscriptstyle F}b_{\rm\scriptscriptstyle J}$}}
\newcommand{\brjhk}{\mbox{$b_{\rm\scriptscriptstyle J}r_{\rm\scriptscriptstyle F}JHK$}}
\newcommand{\jhk}{\mbox{$JHK$}}
\newcommand{\br}{\mbox{$b_{\rm\scriptscriptstyle J}r_{\rm\scriptscriptstyle F}$}}

\newcommand{\mlim}{\mbox{$m_{\rm lim}$}}

\newcommand{\mk}{\mbox{$M_K$}}

\newcommand{\ms}{\mbox{$M^*$}}

\newcommand{\plotone}[1]
    {\centering \leavevmode \psfig{file=#1,width=\plotwidth,clip=}}

\newcommand{\plotfull}[2]
    {\centering \leavevmode \psfig{file=#1,width=#2\fullwidth,clip=}}

\title[6dFGS: Final Redshift Release (DR3) and Large-Scale Structures] 
{The 6dF Galaxy Survey: Final Redshift Release (DR3) and Southern Large-Scale Structures}

\author[Jones et al.]{
\parbox[t]{\textwidth}{
D.\ Heath Jones$^{1}$, 
Mike A.\ Read$^{2}$, 
Will Saunders$^{1}$, 
Matthew Colless$^{1}$,
Tom Jarrett$^{3}$,
Quentin A.\ Parker$^{1,4}$,
Anthony P.\ Fairall$^{5,15}$,
Thomas Mauch$^{6}$, 
Elaine M.\ Sadler$^{7}$
Fred G.\ Watson$^{1}$,
Donna Burton$^{1}$,
Lachlan A.\  Campbell$^{1,8}$,
Paul Cass$^{1}$,
Scott M.\ Croom$^{7}$,
John Dawe$^{1,15}$,
Kristin Fiegert$^{1}$,
Leela Frankcombe$^{8}$,
Malcolm Hartley$^{1}$,
John Huchra$^{9}$,
Dionne James$^{1}$, 
Emma Kirby$^{8}$,
Ofer Lahav$^{10}$, 
John Lucey$^{11}$, 
Gary A.\ Mamon$^{12,13}$,
Lesa Moore$^{7}$,
Bruce A.\ Peterson$^{8}$,
Sayuri Prior$^{8}$,
Dominique Proust$^{13}$,
Ken Russell$^{1}$,
Vicky Safouris$^{8}$,
Ken-ichi Wakamatsu$^{14}$,
Eduard Westra$^{8}$,
and Mary Williams$^{8}$
}
\vspace*{6pt} \\
$^1$Anglo-Australian Observatory, P.O.\ Box 296, Epping, NSW 1710,
Australia {\sf (heath@aao.gov.au)}\\
$^2$Institute for Astronomy, Royal Observatory, Blackford Hill,
Edinburgh, EH9~3HJ, United Kingdom\\
$^3$Infrared Processing and Analysis Center, California Institute of
Technology, Mail Code 100-22,
Pasadena, CA 91125, USA\\
$^4$Department of Physics, Macquarie University, Sydney 2109, Australia\\
$^5$Department of Astronomy, University of Cape Town, Private Bag, 
Rondebosch 7700, South Africa\\
$^6$Astrophysics, Department of Physics, University of Oxford, Keble Road, Oxford, OX1 3RH, UK\\
$^7$School of Physics, University of Sydney, NSW 2006, Australia\\
$^8$Research School of Astronomy \& Astrophysics, The Australian
National University, Weston Creek, ACT 2611, Australia\\
$^9$Harvard-Smithsonian Center for Astrophysics, 60 Garden St  MS20,
Cambridge, MA 02138-1516, USA\\
$^{10}$Department of Physics and Astronomy, University College London,
Gower St, London WC1E 6BT, UK\\ 
$^{11}$Department of Physics, University of Durham, South Road,
Durham DH1~3LE, United Kingdom \\
$^{12}$Institut d'Astrophysique de Paris (CNRS UMR 7095),
98 bis Bd Arago, F-75014 Paris, France\\
$^{13}$GEPI (CNRS UMR 8111), Observatoire de Paris, F-92195 Meudon, France\\
$^{14}$Faculty of Engineering, Gifu University, Gifu 501--1193, Japan\\
$^{15}$deceased.
} \date{Accepted ---. Received ---; in original form ---.}
 
\begin{document}

\maketitle

\begin{abstract}
  We report the final redshift release of the 6dF Galaxy Survey, a combined redshift
  and peculiar velocity survey over the southern sky ($|b|>10^\circ$).
  Its 136\,304 spectra have yielded 110\,256 new extragalactic redshifts 
  and a new catalogue of 125\,071 galaxies making near-complete samples with 
  $(K, H, J, \rf, \bj) \leq (12.65, 12.95, 13.75, 15.60, 16.75)$. The median redshift
  of the survey is 0.053. Survey data, including images, spectra,
  photometry and redshifts, are available through an online database. 
  We describe changes to the information in the database since earlier
  interim data releases. Future releases will include velocity
  dispersions, distances and peculiar velocities for the brightest
  early-type galaxies, comprising about 10\% of the sample. Here we
  provide redshift maps of the southern local universe with $z\leq0.1$,
  showing nearby large-scale structures in hitherto unseen detail. A
  number of regions known previously to have a paucity of galaxies are
  confirmed as significantly underdense regions. The URL of the 6dFGS
  database is http://www-wfau.roe.ac.uk/6dFGS.
\end{abstract}

\begin{keywords}
  surveys --- galaxies: clustering --- galaxies: distances and redshifts
  --- cosmology: observations --- cosmology: large scale structure of
  universe
\end{keywords}

\newpage


\section{Introduction}
\label{sec:introduction}

\begin{table}
\begin{center}
\caption{Comparison of recent wide-area low-redshift galaxy surveys
\label{tab:OtherSurveys}
} 
\begin{tabular}{lccc}
\hline
                                          			& 6dFGS             & 2dFGRS           & SDSS-DR7       \\
\hline
Magnitude limits                          		& $K \leq 12.65$    & $\bj \leq 19.45$ & $r \leq 17.77$   \\
                                          			& $H \leq 12.95$    &                  & (Petrosian)      \\
                                          			& $J \leq 13.75$    &                  &                  \\
                                          			& $\rf \leq 15.60$  &                  &                  \\
                                          			& $ \bj \leq 16.75$ &                  &                  \\
Sky coverage (sr)                         	&  5.2         & 0.5           & 2.86       \\
Fraction of sky                           		& 41\%              & 4\%              & 23\%             \\
Extragalactic sample, $N$                   & 125\,071    & 221\,414   & 644\,951   \\
Median redshift, $z_\frac{1}{2}$          & 0.053        &  0.11        & 0.1        \\
Volume $V$ in $[0.5 z_\frac{1}{2},$  &                   &                  &                  \\
$1.5 z_\frac{1}{2}]$ ~~~~(\cubicMpc)                      & $2.1\times10^{7}$ & $1.7\times10^{7}$ & $7.6\times10^{7}$ \\
Sampling density at $z_\frac{1}{2}$,            &                   &                  &                  \\
$\bar{\rho}=\frac{2N}{3V}$ ~~~~(\invcubicMpc) & $4\times10^{-3}$  & $9\times10^{-3}$ & $6\times10^{-3}$ \\
Fibre aperture (\arcsec)             			& 6.7               & 2.0        & 3.0        \\
Fibre aperture at $z_\frac{1}{2}$    		&                       &                        &                 \\
(\kiloparsec)            					& 4.8               & 2.8        & 3.9       \\

Reference(s)                                 & (1)               & (2)              & (3)              \\
\hline
\end{tabular}\\
\end{center}
\begin{flushleft}
Taking $h = H_0/100$\,\kmsMpc, $\Omega_{M_0} = 0.3$, and $\Omega_{\Lambda_0} = 0.7$.
References: (1)~this paper; (2)~\citet{colless01}, \citet{cole05}; 
(3)~\citet{abazajian09}, www.sdss.org/dr7, (${\tt PetroMag\_r} < 17.77$, ${\tt type} = 3$, 
${\tt zStatus} > 2$, and objects with stellar morphology and $z > 0.001$).
\end{flushleft}
\end{table}

The advent of wide-field multiplexing spectrographs over the past decade
has produced huge advances in our knowledge of the structure and content
of the low-redshift universe. Surveys such as the 2dF Galaxy Redshift
Survey \citep[2dFGRS;][]{colless01} and the Sloan Digital Sky Survey
\citep[SDSS;][]{york00,abazajian09} have characterised the luminosity and clustering
properties of galaxies in unprecedented detail and measured with
unprecedented precision the amount and spatial distribution of dark
matter. They also place tight constraints on $\Lambda$-Cold Dark Matter
models of the universe \citep[e.g.][]{spergel07} when combined with the
results of supernovae distance measurements \citep{schmidt98, riess98,
perlmutter99} and the cosmic microwave background \citep{bennett03}.
Within this context, the focus has shifted towards an improved
understanding of galaxy mass assembly and structure formation generally
\citep[e.g.][]{baugh06}. A combined redshift and peculiar velocity
survey, with dynamical measures of galaxy masses and large-scale
motions, offers even better constraints on parameters of cosmological
interest than a survey of redshifts alone \citep{burkey04, zaroubi05}.

The 6dF Galaxy Survey\footnote{6dFGS home: http://www.aao.gov.au/6dFGS}
\citep[6dFGS;][]{jones04,jones05} is a near-infrared-selected redshift
and peculiar velocity survey that is complete to total extrapolated magnitude limits
$(K,H,J)=(12.65,12.95,13.75)$\footnote{These limits
differ slightly from those reported in \citet{jones04} and
\citet{jones05} due to subsequent revision of the input magnitudes by
2MASS and SuperCOSMOS; see Sec.~2.}
over 83~percent of the southern sky.  The near-infrared total $JHK$
magnitudes are taken from the Two Micron All-Sky Survey (2MASS) Extended
Source Catalog \citep[XSC;][]{jarrett00}, while \bj\rf\ photometry comes
from SuperCOSMOS \citep{hambly01a,hambly01b}, 
following its recalibration for the 2dF Galaxy
Redshift Survey \citep{cole05}.

Near-infrared (NIR) selection of the primary galaxy samples is
advantageous because it closely tracks the older stellar populations
that dominate the stellar mass in galaxies. Furthermore, extinction (both 
internal and Galactic) is greatly lessened and stellar mass-to-light ratios are more tightly
defined \citep{bell01}. Two secondary optically-selected
samples are complete to $(\rf,\bj)= (15.60,16.75)^{2}$. A number
of smaller samples, selected from various catalogues and wavelengths,
fill out the target allocations. The peculiar velocity survey uses
velocity dispersions and photometric scale-lengths to derive dynamical
masses and Fundamental Plane distances and peculiar velocities for a
subset of more than 10\,000 bright, early-type galaxies. 

The 6dFGS magnitude limits are $\sim 1.5$~mag brighter than the
magnitudes at which incompleteness starts to affect the 2MASS XSC
($\kb \lesssim14$). Examination of the bivariate distribution of surface
brightness and galaxy luminosity for the entire 6dFGS sample (Jones
et~al., in prep) shows sample selection to be robust against surface
brightness selection effects \citep[see e.g.][]{bell03,mcintosh06}. 
The limiting isophote at which 6dFGS magnitudes were measured 
($\mu_K = 20$\,mag\,arcsec$^{-2}$) is brighter than the values at which 
2MASS was found to be incomplete by these authors.

The 6dFGS has thus far been used in studies of large-scale structure
\citep[e.g.][]{fleenor05, fleenor06, boue07, radburnsmith06,proust06},
luminosity and stellar mass functions \citep[][Jones et~al.,
in prep]{jones06}, the influence of local density and velocity
distributions \citep{erdogdu06a, erdogdu06b}, among others. 
The Early and First Data Releases (see below) alone yielded new 
redshifts for 277 ACO clusters ($z \lesssim 0.1$) without previous redshifts
\citep{andernach05}, and the full data have yielded more than 400.
Examples of the full 3-d space structure of the 6dFGS can be seen
in \citet{fluke09}. The 6dFGS has also been 
used to study special interest samples selected for their luminosity at
x-ray and radio wavelengths \citep{sadler06,
mauduit07, mauch07}. Future surveys with next
generation radio telescopes such as ASKAP and the SKA 
\citep[e.g.][]{blake04,vandriel05,rawlings06}
will also benefit from the legacy of 6dFGS, as they probe comparable volumes
in H{\sc I} with the benefit of prior redshift information across most
of the southern sky.

This paper describes the final data release of 6dFGS redshifts. Earlier
incremental data releases in 2002 December, 2004 March and 2005 May have
made the first 90k redshifts publicly available through an online
database. In Sec.~2 we give an overview of the 6dF Galaxy Survey
including the characteristics and scope of the data set. Section~3 describes
the final instalment as well as its access through our online database.
Details of changes and additions superceding earlier releases are also
given. In Sec.~4 we present redshift maps of the southern sky in both
equatorial and Galactic coordinate projections, and discuss major large-scale 
structures. Concluding remarks are made in Sec.~5.


\section{Survey Overview}
\label{sec:overview}

\subsection{Background}

\begin{figure*}

\plotfull{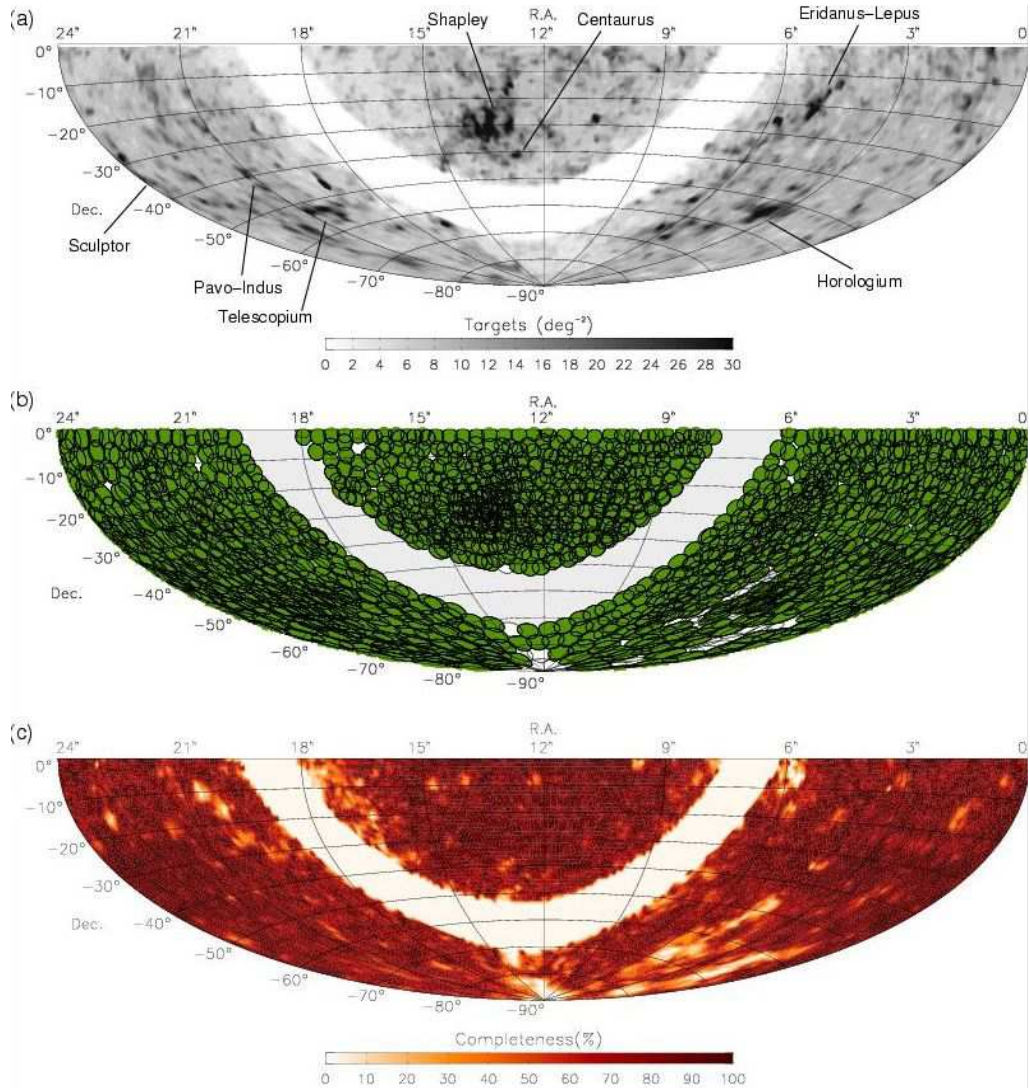}{0.8}

\caption{(a)~Density of 6dFGS target sources (per square degree) on the
  sky; key supercluster over-densities are labelled. (b)~Full 6dFGS
  field coverage (filled discs) and unobserved target fields (open
  circles). (c)~Redshift completeness for $\kb \leq 12.65$. All panels show
  equal-area Aitoff projections.}
\label{fig:skycoverage}
\end{figure*}

The primary references for detailed information about the 6dFGS are
\citet{jones04} and this paper. The former describes target selection and field allocations,
the 6dF instrument, and data reduction and redshifting methodology. It
also characterises the First Data Release (DR1; 46k redshifts) and the
online database. \citet{jones05} describes the Second Data Release (DR2;
83k redshifts) and discusses a number of small changes to the data.
Earlier papers describe the 6dF instrument \citep{parker98,watson00} and
the field placement algorithm used to optimise target coverage
\citep{campbell04}. Database users are encouraged to consult these and
other papers on the 6dFGS Publications web
page\footnote{http://www.aao.gov.au/6dFGS/Publications}. This
paper marks the final public data release of 6dFGS redshift data.

The observations for this survey were carried out using the Six Degree Field (6dF) fibre-fed
multi-object spectrograph at the UK Schmidt Telescope (UKST) over
2001 May to 2006 January \citep{jones04}. Target fields covered the
$\sim17\,000$\sqdeg\ of southern sky more than $10^\circ$ from the
Galactic Plane\footnote{The \bj\ and \rf\
surveys of 6dFGS are limited to $|b|>20^\circ$ in order to mitigate
the effect of higher Galactic extinction in the optical at lower latitudes.},
approximately ten times the area of the 2dF Galaxy Redshift Survey
\citep[2dFGRS;][]{colless01} and more than twice the spectroscopic areal coverage
of the Sloan Digital Sky Survey seventh data release \citep[SDSS DR7;][]{abazajian09}. 
Table~\ref{tab:OtherSurveys} shows a
comparison of the 6dFGS to these two major surveys.  In terms
of secure redshifts, 6dFGS has around half the number of 2dFGRS and
one-sixth those of SDSS DR7 ($r < 17.77$). The co-moving volume covered of 6dFGS is
about the same as 2dFGRS at their respective median redshifts, and
around 30~percent that of SDSS DR7. In terms of fibre aperture size, the
larger apertures of 6dFGS (6.7\arcsec) give a projected diameter of
4.8~\kiloparsec\ at the median redshift of the survey, covering 40~percent more
projected area than SDSS at its median redshift, and more than three
times the area of 2dFGRS. By any measure, the scale of 6dFGS is readily
comparable to those of SDSS and 2dFGRS, and like those surveys, its
legacy is a permanent public database, which is unique in its
scope, depth and southern aspect.

Figure~\ref{fig:skycoverage} shows the sky distribution of 6dFGS targets
and fields. Of the 1\,526 fields observed, 1\,447 contributed
data to the final survey. The remaining 5 percent were rejected
for reasons of quality control. Of the 1\,447, around half have
completeness greater than 90 percent, and more than two thirds
have completeness greater than 85 percent. Although most sky
regions are effectively covered twice, around 50 fields near the LMC and
the South Pole were not observed by the conclusion of the survey.
Sky redshift completeness (Fig.~\ref{fig:skycoverage}$c$)
is generally high (85 percent or greater) but diminishes in regions with
insufficient coverage or affected by poor conditions. In terms of survey
limiting magnitude, \mlim, completeness is greater than 85 percent for
$m < (\mlim - 0.75)$ in fields with completeness 90 percent or higher,
and for $m < (\mlim - 2)$ in fields with completenesses of 70 to 80 percent.


\subsection{Redshift Distribution}

\begin{figure}

\plotone{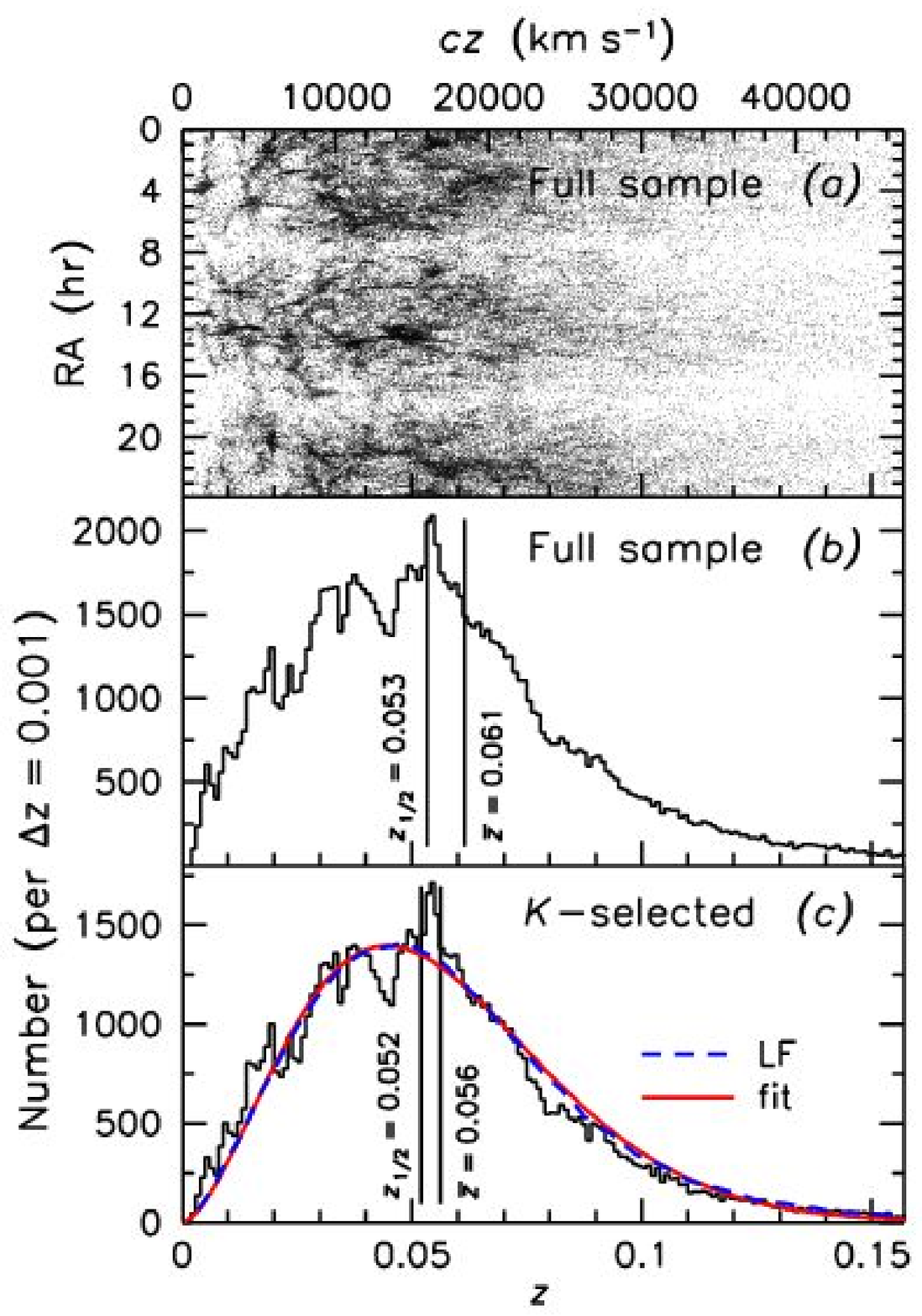}

\caption{Distribution of all 6dFGS redshifts in terms of (a)~right ascension
and (b)~number. Panel
(c)~shows the same as (b) but limited to the primary \kb-selected sample.
The dashed blue line is the redshift distribution calculated from the 
\kb-band luminosity function of the same sample (Jones et al., in prep).
The solid red line is an empirical fit to the blue curve.}
\label{fig:nz}
\end{figure}

The median redshift for 6dFGS is $z_{1/2} = 0.053$, roughly half that of SDSS 
and 2dFGRS, and twice that of the 2MASS Redshift Survey \citep[2MRS;][]{erdogdu06a}.
Figure~\ref{fig:nz} shows the number distribution of 6dFGS redshifts
for both the full sample, $N(z)$, (panel $b$; 125\,071 sources) as well as the $K$-selected
primary targets, $N_K(z)$, (panel $c$; 93\,361). Both samples show the skewed
distributions typical for magnitude-limited surveys, which is accentuated in Fig.~\ref{fig:nz}($b$)
by the inclusion of Additional Target samples that stretch the overall distribution
to higher redshifts. This is also reflected in their interquartile ranges: $[0.034,0.074]$ for the full
sample, compared to the slightly narrower $[0.034,0.070]$ for the \kb-selected sample.
The localised peaks in $N(z)$ and $N_K(z)$ are due to individual large-scale structures,
clearly seen in Fig.~\ref{fig:nz}($a$) when redshifts are spread across RA.
The gaps in ($a$) centred on R.A. 8\,hr and 17\,hr correspond to
the unsurveyed regions around the Galactic Plane.

The limit of the \kb-selected sample ($\kb \leq 12.65$) encompasses galaxies with
luminosities $\mk \leq -23.24$ at the median redshift ($z_{1/2} = 0.053$), around
0.6\,mag fainter than \ms, the characteristic turn-over point  in the \kb-band
luminosity function from the same sample \citep[][Jones et al, in prep.]{jones06}. 
Integrating this luminosity function over the volume covered by 6dFGS in each
redshift shell $\Delta z$ yields the expected number-redshift distribution $N_{\rm LF}(z)$.
As the \kb-band luminosity function contains completeness corrections that the
raw $N_K(z)$ distribution does not, the ratio
\begin{equation}
\frac{\int_0^\infty\,N_K(z)\,dz}{\int_0^\infty\,N_{\rm LF}(z)\,dz} = 0.9245
\end{equation}
is slightly less than unity. The blue dashed curve representing $N_{\rm LF}(z)$
in Fig.~\ref{fig:nz}($c$) has been scaled by this amount, and the ratio of the two
distributions $N_K(z)\,/\,N_{\rm LF}(z)$ gives the normalisation due to overall
incompleteness. Any redshift differences between the curve and the data are due
to magnitude-dependent incompleteness, which in turn imparts redshift differences
in selection. Furthermore, a Schechter function is not a perfect fit to the luminosity
function across all luminosities. The reader is referred to the 6dFGS luminosity function papers
\citep[][Jones et al.\ in prep.]{jones06} for a more detailed discussion of survey
selection functions.

The curve $N_{\rm LF}(z)$ (uncorrected for incompleteness)
is well-fit by the empirical function
\begin{equation}
N_{\rm fit}(z) = A\,z^\gamma\,\exp[-( z / z_{\rm p})^\gamma]
\end{equation}
with values $\gamma = 1.6154 \pm 0.0001$, 
$z_{\rm p} = 0.0446 \pm 0.0001$,
and
$A = 622978 \pm 10$ (Fig.~\ref{fig:nz}$c$; red solid line, also scaled by 0.9245). 
This 3-parameter function is a simpler variant of the 4-parameter fits used by
\citet{erdogdu06a} and \citet{colless01} for the 2MRS and 2dFGRS samples,
respectively, but fits the 6dFGS \kb-band sample well.
In this case, the value of $z_{\rm p}$ locates the peak in the distribution, 
which is slightly lower than the median of the data, and corresponds to a
limiting absolute magnitude of $\mk = -22.97$ ($\sim 1$~mag fainter than $\ms$).
Even so, the remarkable consistency between the \kb-band luminosity function 
distribution ($N_{\rm LF}(z)$) and that of the data ($N_K(z)$) underscores the 
homogeneity of the primary sample.


\subsection{Sample Composition}

The original target catalogue for 6dFGS contained 179\,262
sources, one third of which originated from outside the near-infrared
selected catalogues. Around 8 percent of these sources had existing
redshifts from ZCAT \citep[9\,042;][]{huchra02}, the 2dF Galaxy
Redshift Survey \citep[5\,210;][]{colless01}, or the Sloan
Digital Sky Survey DR7 \citep[563;][]{abazajian09}. 6dFGS spectra were
obtained in 136\,304 source observations and yielded 126\,754
unique redshifts of varying quality. 

6dFGS redshift quality, $Q$, was classified on a scale of 1 to 6 through
visual assessment of every redshift, with $Q=1$ assigned to unusable
measurements, $Q=2$ to possible but unlikely redshifts, $Q=3$ for reliable 
redshifts and $Q=4$ for high-quality redshifts. Stars and other confirmed Galactic
sources are assigned $Q=6$ (there is no $Q=5$). 
Some legitimate QSO redshifts classified earlier in the survey carry 
$Q=2$ because no QSO-specific templates were employed to classify QSO 
redshifts until mid-way through the survey.
Unlike SDSS, no lower velocity limit has been used to 
discriminate between Galactic and extragalactic sources; assignment
of $Q=6$ is on the basis of spectral appearance as well as
recession velocity. Cases of overlap between galaxies and foreground stars
evident from imaging data were re-examined, and are discussed in Sec.~\ref{sec:release}.
Table~\ref{tab:distribution} gives the breakdown of these numbers across
individual 6dFGS sub-samples.

\begin{table}
\begin{center}
\caption{Breakdown of 6dFGS and literature redshifts.
\label{tab:distribution}
}
\begin{tabular}{lr}
\hline\\
{\bf 6dFGS by $Q$ value:} & \\
$Q=1$, unusable data   &    5\,787 ( 4.6 pc) \\
$Q=2$, unlikely redshifts   &    5\,592 ( 4.4 pc) \\
$Q=3$, reliable redshifts   &    8\,173 ( 6.4 pc) \\
$Q=4$, high quality redshifts   &  102\,083 (80.5 pc) \\
$Q=6$, Galactic sources   &    5\,119 ( 4.0 pc) \\
{\bf Total}   &  126\,754    (100 pc) \\
& \\
{\bf Literature redshifts:} & \\
SDSS        &     563 ( 3.8 pc) \\
2dFGRS   &    5\,210 (35.2 pc) \\
ZCAT         &    9\,042 (61.0 pc) \\
{\bf Total}   &    14\,815 (100 pc)  \\
& \\
\hline\\

\end{tabular}
\end{center}
{\bf References:}
SDSS: \citet{abazajian09}, 2dFGRS: \citet{colless01}, ZCAT: \citet{huchra02}\\
\end{table}

\begin{table}
\begin{center}
\caption{Final numbers of spectra and redshifts in the 6dFGS samples.
\label{tab:breakdown}
}
\begin{tabular}{rlrrrr}
\hline
   ID & Survey sample          & 6dFGS   & Good & Lit.\ & Total \\
      &                          & spectra & $z$     & $z$   & $z$   \\
\hline
   1  & 2MASS $K_s \leq 12.75$            &  97\,020 &  83\,995 &   9\,340 &  93\,335 \\
   3  & 2MASS $H  \leq 12.95$               &   2\,021 &   1\,742   &    255     &   1\,997 \\
   4  & 2MASS $J  \leq 13.75$                &   1\,284 &   1\,096    &    175    &   1\,271 \\
   5  & DENIS $J  \leq 14.00$                 &    629    &    488       &    115     &    603 \\
   6  & DENIS $I \leq 14.85$                   &    504    &    234       &    109     &    343 \\
   7  & {\small SuperCOS} $\rf<15.6$    &   5\,773 &   5\,025    &   1\,221 &   6\,246 \\
   8  & {\small SuperCOS} $\bj<16.75$  &   6\,516 &   5\,885   &   1\,236 &   7\,121 \\
  78  & Dur./UKST extension                   &    271    &    207       &     30      &    237 \\
  90  & Shapley supercluster                    &    630   &    494       &     40      &    534 \\
 109  & Horologium sample                     &    469   &    384       &     41      &    425\\
 113  & {\small ROSAT} all-sky survey  &   1\,961 &   1\,126 &    190      &   1\,316 \\
 116  & 2MASS Red AGN                        &   1\,141 &    438      &    140      &    578 \\
 119  & HIPASS ($>4\sigma$)                &    439    &    354       &    116     &    470 \\
 125  & SUMSS/NVSS radio                   &   2\,978 &   1\,351   &    272      &   1\,623 \\
 126  & IRAS FSC ($>6\sigma$)             &   5\,994 &   4\,208   &   1\,239   &   5\,447 \\
 129  & Hamburg-ESO QSOs                  &   2\,006 &    624      &    123       &    747 \\
 130  & NRAO-VLA QSOs                        &   2\,673 &    293      &     41         &    334 \\
 $\geq999$  & unassigned targets$^\dagger$    &   3\,995 &   2312       &     132         &    2\,444 \\
      & {\bf Total}                                           & 136\,304 & 110\,256 &  14\,815 & 125\,071 \\
      \hline
\end{tabular}
\end{center}
$\dagger$~~Objects removed from the initial target list (due to changes in\\
\hspace*{12pt}the 2MASS source catalogue after 6dFGS was underway).\\
\hspace*{12pt}ID $= 999$ or 9999 in these cases.\vspace{2mm}\\
{\bf Columns:}\\
(1) ID: Programme ID ({\tt PROGID} in the database; see Sec.~\ref{sec:release}).\\
(2) Survey sample: first sample (in order of {\tt PROGID}) in which\\
\hspace*{12pt} object is found.\\
(3) 6dFGS spectra: number of spectra obtained for this sample. \\
\hspace*{12pt}Note that some objects were observed more than once. The\\
\hspace*{12pt}numbers include spectra of all qualities and Galactic  sources.\\
(4) Good $z$: number of robust extragalactic 6dFGS redshifts,\\
\hspace*{12pt} (those with $Q = 3$ or 4). Reflects contents of database.\\
(5) Lit.\ $z$: additional literature extragalactic redshifts (ignoring\\
\hspace*{12pt} repeats and overlaps).\\
(6) Total $z$: total number of extragalactic redshifts for objects\\
\hspace*{12pt} in this sample.\\
\end{table}

Only $Q=3,4$ redshifts should be used in any galaxy analysis.
(The distinction between $Q=3$ and $Q=4$ is less important than that
between $Q=2$ and $Q=3$, since the former represent a successful
redshift in either case.) Galaxies with repeat observations have all spectra retained 
in the database, and the final catalogued redshift is a weighted mean of the 
measurements with $Q=3,4$, excluding redshift blunders.
Descriptions of the redshift quality scheme in its previous forms can be
found in Sec.~2.1 of \citet{jones05} and Sec.~4.4 of \citet{jones04}.

Unreliable ($Q=2$) or unusable ($Q=1$) galaxy redshifts together comprise
around  8 percent of the redshift sample. Galactic sources ($Q=6$)
represent another 4 percent. The remaining
110\,256 sources with $Q=3,4$ are the robust extragalactic 6dFGS
redshifts that should be used (alongside the 14\,815 literature redshifts)
in any analysis or other application.
Tables~\ref{tab:distribution} and \ref{tab:breakdown} give the breakdown of 
these numbers across various 6dFGS sub-samples.

\begin{figure}

\plotone{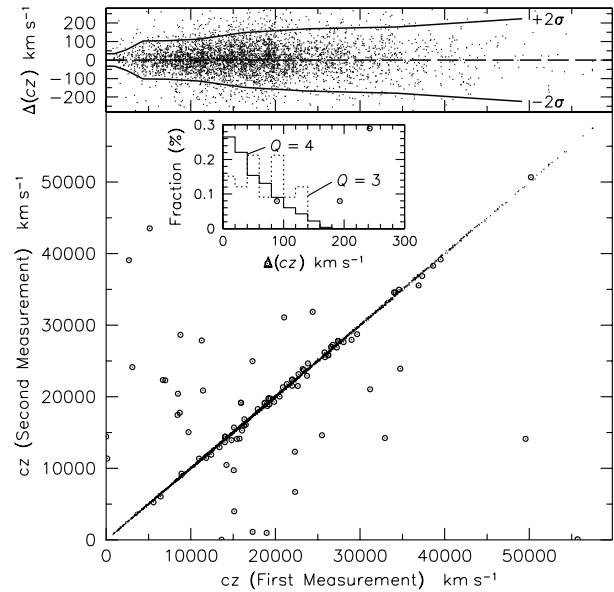}

\caption{(Top panel:) Repeat 6dF redshift measurements for a sample of
6dFGS galaxies obtained with the VPH gratings over the period 2002.5 to 2006
(4570 galaxies). Redshift blunders (circled) are
those for which $|\,\Delta cz\,| > 330$\kms. (Inset panel:) Distribution of the $|\,\Delta cz\,|$ 
diffferences for the individual redshift quality $Q=3$ (dotted line) 
and $Q=4$ (solid line) samples, normalised 
to the total sample size in each case. 
(Bottom panel:) Distribution of redshift difference as a function of 
redshift, with a running $\pm2\sigma$ boundary (solid lines).}
\label{fig:6dfVS6df}
\end{figure}


\subsection{Redshift uncertainties and blunder rates}

Redshift uncertainties and blunder rates were estimated from the sample
of 6dFGS galaxies with repeat redshift measurements. There were
8\,028 such redshift pairs,  43\,percent of which were first-year (pre-2002.5) data.
Most repeat measurements were made because of a low-quality
initial measurement, or because of a change in the field tiling strategy after the first
year of observations. We define a blunder as a redshift mismatch of more than 330\,\kms 
($5\sigma$) between a pair of redshift measurements that we would expect to agree. 
The blunder rate on individual 6dFGS redshifts is 1.6\,percent,
the same as reported for the First Data Release \citep{jones04}.
In late 2002, a new transmissive Volume-Phase Holographic (VPH)
gratings (580V and 425R) replaced the existing reflection gratings (600V and 316R),
resulting in improved throughput, uniformity, and data quality. Excluding first-year
repeats reduces the individual blunder rate to 1.2\,percent. 
While the first-year data represent nearly half of
all repeat measurements, they represent less than a fifth of the overall survey.
Table~\ref{tab:blunder} summarises the blunder rates and other statistics for both
the full and post-first-year data.

Figure~\ref{fig:6dfVS6df} shows repeat redshift measurements for 6dFGS observations
with the VPH gratings, representative of the great majority of survey spectra
(around 80 percent: 4\,570 sources spanning 2002.5 to 2006). Blunder measurements
(106 of them) have been circled, and the scatter in redshift offset, $\Delta cz$, as a function of redshift
is also shown. Not surprisingly, redshifts becoming increasingly difficult to secure as one moves
to higher values. The inset in Figure~\ref{fig:6dfVS6df}
displays the distribution in $|\,\Delta cz\,|$ for measurement pairs grouped by
their redshift quality $Q=3$ or 4 classifications. There are 3\,611 pairs in the
non-blunder sample with both measurements of quality $Q=4$, with scatter implying
a redshift uncertainty in an individual $Q=4$ measurement of $\Delta cz(4) = 46$\kms.
Likewise, the scatter in the much smaller $Q=3$ sample (33 pairs) suggests
$\Delta cz(3) = 55$\kms. If we include the pre-2002.5 non-VPH data, the implied
redshift uncertainties are unchanged for $Q=4$ and increase slightly in the
case of $Q=3$ ($\Delta cz(3) = 67$\kms). Note that these redshift uncertainties
are less than those estimated in \citet{jones04} from the First Data Release,
demonstrating the improved integrity of the 6dFGS data since the early releases.

\begin{table}
\begin{center}
\caption{Redshift uncertainties and blunder rates from both internal and external comparisons of 6dFGS. 
\label{tab:blunder}
} 
\vspace{6pt}
\begin{tabular}{llr}
\hline \\

 {\bf 6dFGS (full sample):}      &  \\
 Total repeat measurements ($Q \geq 3$): & 			8028 \\
 RMS scatter of all redshift measurement pairs$^\dagger$ & $66$\kms \\
 $Q=4$ redshift uncertainty (6051 sources) & 			$45$\kms \\
 $Q=3$ redshift uncertainty (104 sources)     &  		$67$\kms\\
    &  \\
 Number of blunders$^\ddagger$ ($Q\geq 3$): & 		260\\
 6dFGS pair-wise blunder rate: & 					3.2\% \\
 6dFGS single-measurement blunder rate: & 			1.6\% \\
     \\
  {\bf 6dFGS (VPH grating only, 2002.5 -- 2006):}      &  \\
 Total repeat measurements ($Q \geq 3$): & 			4570 \\
 RMS scatter of all redshift measurement pairs$^\dagger$ & $67$\kms \\
 $Q=4$ redshift uncertainty (3611 sources) & 			$46$\kms \\
 $Q=3$ redshift uncertainty (33 sources)     &  			$55$\kms\\
      &  \\
 Number of blunders$^\ddagger$ ($Q\geq 3$): & 		106 \\
 6dFGS pair-wise blunder rate: & 					2.3\% \\
 6dFGS single-measurement blunder rate: & 			1.2\% \\
  
   \hline\\
 {\bf 6dFGS (VPH only) vs. SDSS DR7:}    & \\
 Number of comparison sources ($Q\geq3$):  &  		2459 \\
 Number of blunders$^\dagger$ ($Q\geq3$): & 		95 \\
 Pair-wise blunder rate: &  						3.9\% \\
 Implied blunder rate for SDSS:    &     				2.7\% \\                 
\hline \\
\end{tabular}
\end{center}
\begin{flushleft}
$\dagger$~~Clipping the most extreme $10$\% of outliers ($5$\% either side).\\
$\ddagger$~~A blunder is defined as having $\Delta cz > 330$\kms ($5\sigma$).\\
\end{flushleft}
\end{table}

\begin{figure}

\plotone{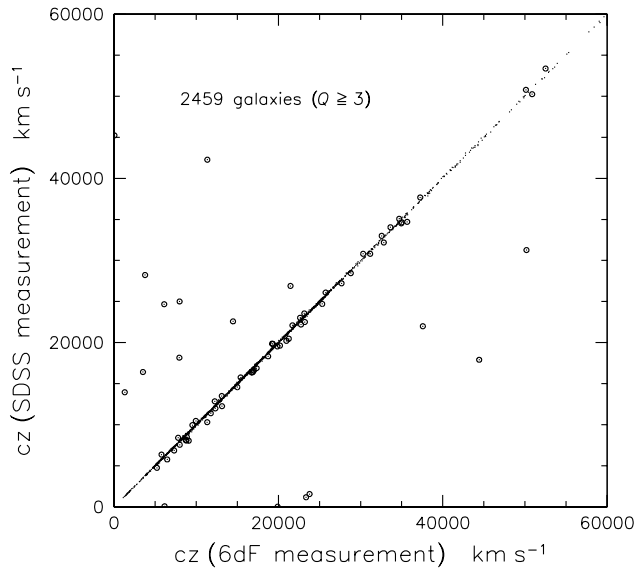}

\caption{Redshift comparison of 6dFGS (VPH grating) with SDSS Data Release 7
\citep{abazajian09}.}
\label{fig:litVS6df}
\end{figure}

An external comparison of 6dFGS redshifts to those overlapping the Seventh Data
Release (DR7) of the Sloan Digital Sky Survey \citep[SDSS;][]{abazajian09} was also
made and is shown in Fig.~\ref{fig:litVS6df}. Although the full SDSS DR7 contains
over a million classified extragalactic spectra, almost all are too northerly to overlap
significantly with the southern 6dFGS or are too faint. However, the 2\,459 sources in
common to both catalogues provide a valuable test of redshift success rates.
The pair-wise blunder fraction in this case is 3.9\,percent. Splitting this with the 
6dFGS blunder rate of 1.2\,percent implies an SDSS blunder rate of 2.7\,percent,
although the 6dFGS blunder rate at the fainter SDSS magnitudes is likely to be somewhat
higher than the 1.2\,percent measured overall.

\section{New Data Release}
\label{sec:release}

\subsection{Online database}

\begin{figure*}

\plotfull{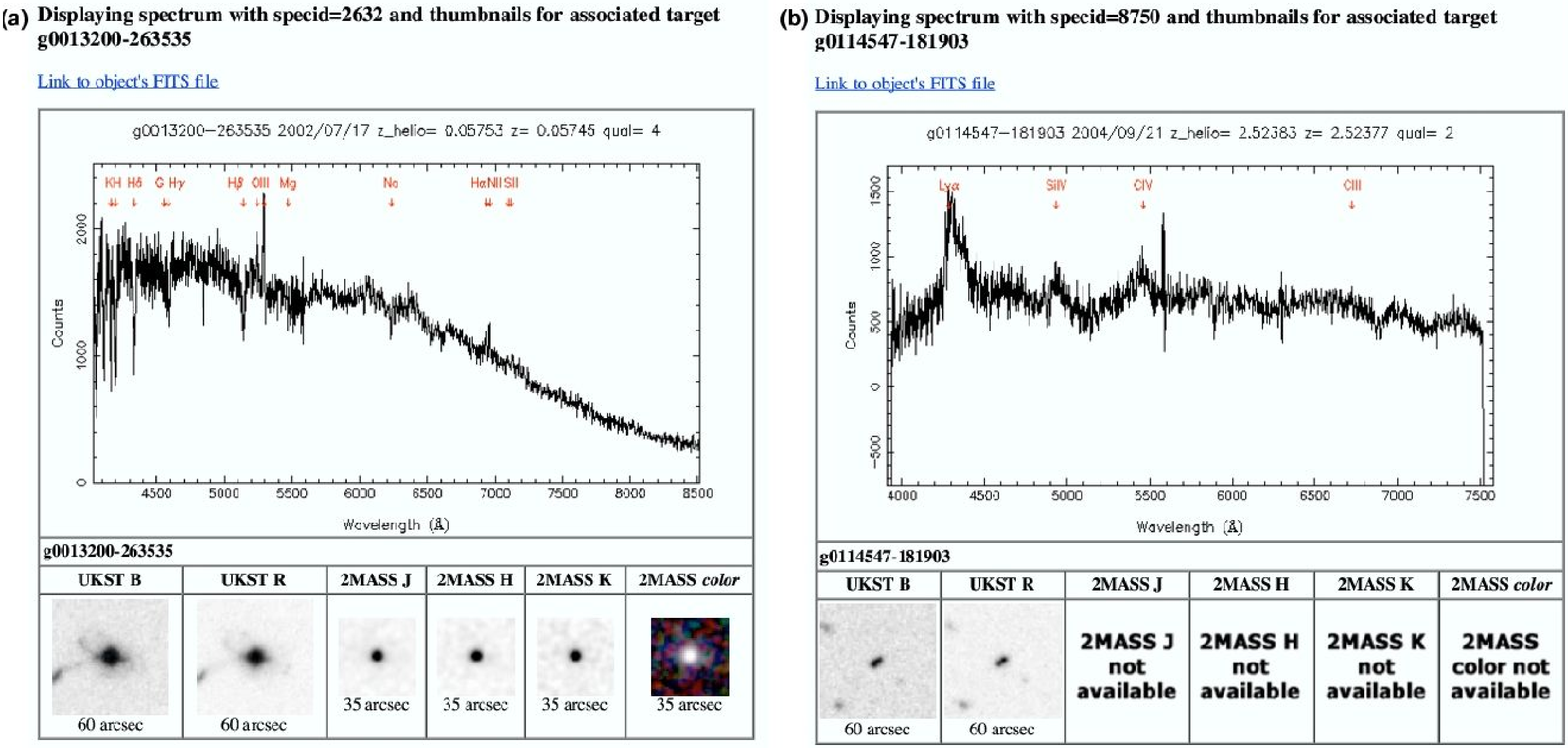}{1.0}

\caption{Example spectroscopic and photometric frames from the 6dFGS
  online database for (a)~a nearby bright galaxy at $z=0.057$ ($Q=4$)from the $K$-selected
  sample ({\tt PROGID} = 1), and (b)~a candidate double QSO at $z=2.524$ ($Q=2$) from the
  Hamburg-ESO QSO sample ({\tt PROGID} = 129). 2MASS and UKST frames are only
  available for sources selected as part of the original 6dFGS primary
  samples, where available in one or more of \khjrb.}
\label{fig:example}
\end{figure*}

The 6dFGS Online Database is hosted 
at the Wide Field Astronomy Unit of the Institute for
Astronomy\footnote{http://www-wfau.roe.ac.uk/6dFGS} at the University of
Edinburgh. Data are grouped into 15 inter-linked tables consisting of
the master target list, all input catalogues, and their photometry.
Users can obtain FITS and JPEG files of 6dFGS spectra as well as 2MASS and
SuperCOSMOS postage stamp images in \jhk\ and \br\ where available, and
a plethora of tabulated values for observational quantities and derived
photometric and spectroscopic properties. The database can be queried in
either its native Structured Query Language (SQL) or via an HTML
web-form interface. More complete descriptions are given elsewhere
\citep{jones04,jones05}, although several new aspects of the database
are discussed below.  Figure~\ref{fig:example} shows two examples of the
way data are presented in the database.

Table~\ref{tab:paramContents} shows the full parameter listing for the
6dFGS database. Individual database parameters are grouped into lists of
related data called {\sl tables}. Parameter definitions are given in
documentation on the database web site. The {\tt TARGET} table contains
the original target list for 6dFGS, and so contains both observed and
unobserved objects. Individual entries in this table are celestial sources, and the 
{\tt TARGETID} parameters are their unique integer identifiers.
Note that the original target list {\em cannot} be
used to estimate completeness, due to magnitude revisions in both the
2MASS XSC and SuperCOSMOS magnitudes subsequent to its compilation.
Item (iv) below discusses this important issue in more detail.

The {\tt SPECTRA} table holds the redshift and other spectroscopic data
obtained by the 6dF instrument through the course of the 6dFGS. Many new
parameters have been introduced to this table for this release
(indicated in Table~\ref{tab:paramContents} by the $\dagger$ symbol).
Individual entries in this table are spectroscopic observations, meaning
that there can be multiple entries for a given object. The {\tt SPECID}
parameter is the unique integer identifier for 6dFGS observations.

Most 6dFGS spectra consist of two halves, observed separately through
different gratings, and subsequently spliced together: a V portion
($\lambda\lambda3900$--5600\AA) and an R portion
($\lambda\lambda5400$--7500\AA).\footnote{V and R here are not related
to standard $V$ or $R$ passbands.} (Data taken prior to October 2002
used different gratings, spanning 4000--5600\AA\ and 5500--8400\AA.)
Various parameters in {\tt SPECTRA} belonging to the individual V or R
observations carry a {\tt \_V} or {\tt \_R} suffix, and are listed in
Table~\ref{tab:paramContents} for V (with slanted font to indicated that
there is a matching set of R parameters).

The {\tt TWOMASS} and {\tt SUPERCOS} tables hold relevant 2MASS XSC and
SuperCOSMOS photometric and spatial information. Likewise, the remaining
eleven tables contain related observables from the input lists
contributing additional 6dFGS targets to {\tt TARGET}.
While some of the parameter names have been duplicated between tables
(e.g.\ {\tt MAG\_1, MAG\_2}) their meaning changes from one table to the next,
as indicated in Table~\ref{tab:paramContents}.

Database tables can be queried individually or in pairs. Alternatively,
positional cross-matching (R.A.\ and Dec.) can be done between database
sources and those in a user-supplied list uploaded to the site. Search
results can be returned as HTML-formatted tables, with each entry
linking to individual GIF frames showing the 6dFGS spectrum alongside
its \bj\rf$JHK$ postage stamp images, as shown in
Fig.~\ref{fig:example}. Individual object FITS files of the same data
can also be accessed in this way. Long database returns can also be
emailed to the user as an ASCII comma-separated variable (CSV) text
file. Alternatively, the FITS files of all objects found through a
search can be emailed to the user as a single tar file under a 
{\sl TAR saveset} option.

\begin{table*}
\begin{center}
\caption{Full parameter listing for all tables in the 6dFGS database.
\label{tab:paramContents}
} 
\begin{tabular}{llcl}
\hline \\
Table name & Description & {\tt PROGID} & Parameters \\
\hline \\
{\tt TARGET} & the master target list & $-$ & {\tt TARGETID, TARGETNAME, HTMID, RA, DEC, CX, CY, CZ, GL, GB, } \\
 & & & {\tt A\_V, PROGID, BMAG, RMAG, SG, ZCATVEL, ZCATERR, ZCATREF, } \\
 & & & {\tt BMAGSEL, RMAGSEL, TEMPLATECODE$\dagger$, FRAMENAME} \\
{\tt SPECTRA} & redshifts and observational data & $-$ & {\tt SPECID, TARGETID, TARGETNAME, OBSRA, OBSDEC, MATCH\_DR, } \\
 & & & {\tt HTMID, CX, CY, CZ, Z\_ORIGIN, Z, Z\_HELIO, QUALITY, ABEMMA,} \\
 & & & {\tt NMBEST, NGOOD, Z\_EMI, Q\_Z\_EMI, KBESTR, R\_CRCOR, Z\_ABS, } \\
 & & & {\tt Q\_Z\_ABS, Q\_FINAL, IALTER, Z\_COMM, ZEMIBESTERR, ZABSBESTERR,}\\
 & & & {\tt ZFINALERR,}  {\sl TITLE\_V, CENRA\_V, GRATSLOT\_V, CENDEC\_V, } \\
 & & & {\sl APPRA\_V, APPDEC\_V, ACTMJD\_V, CONMJD\_V, PROGID\_V,} \\
 & & & {\sl  LABEL\_V, OBSID\_V, RUN\_V, EXP\_V, NCOMB\_V, GRATID\_V,} \\
 & & & {\sl  GRATSET\_V, GRATBLAZ\_V, SOURCE\_V, FOCUS\_V, TFOCUS\_V, } \\
 & & & {\sl GAIN\_V, NOISE\_V, CCD\_V, UTDATE\_V, UTSTRT\_V,}\\
 & & & {\sl MJDOBS\_V, NAME\_V, THPUT\_V, RA\_V, DEC\_V, X\_V, Y\_V, } \\
 & & & {\sl XERR\_V, YERR\_V, THETA\_V, FIBRE\_V, PIVOT\_V, RECMAG\_V, } \\
 & & & {\sl PID\_V,} {\tt FRAMENAME,} {\sl AXISSTART\_V, AXISEND\_V,} {\tt MATCHSPECID,}\\
 & & & {\tt Z\_INITIAL$\dagger$, Z\_HELIO\_INITIAL$\dagger$, Z\_UPDATE\_FLAG$\dagger$, Z\_UPDATE\_COMM$\dagger$,} \\
 & & & {\tt SLIT\_VANE\_CORR$\dagger$, QUALITY\_INITIAL, XTALKFLAG$\dagger$, XTALKSCORE$\dagger$, }\\
 & & & {\tt XTALKVELOFF$\dagger$, XTALKCOMM$\dagger$, QUALITY\_UPDATE\_COMM, DEPRECATED$\dagger$} \\
 & & & {\tt REVTEMPLATE$\dagger$, REVCOMMENT$\dagger$, Z\_COMM\_INITAL$\dagger$} \\
 {\tt TWOMASS} & \jhk\ 2MASS input catalogues & 1 ($K$), & {\tt OBJID, CATNAME, TARGETNAME, TARGETID, RA, DEC, PRIORITY,} \\
& & 3 ($H$), & {\tt MAG\_1}, {\tt PROGID, MAG\_2}, {\tt J\_M\_K20FE, H\_M\_K20FE,} \\
& & 4 ($J$) & {\tt K\_M\_K20FE, RADIUS, A\_B, MUK20FE, CORR, J, H, KEXT, K,} \\
& & & {\tt KEXT\_K, PREVCATNAME$\dagger$, RTOT$\dagger$, JTOT$\dagger$, HTOT$\dagger$, KTOT$\dagger$}\\
{\tt SUPERCOS} & \br\ SuperCOSMOS input catalogues & 8 (\bj), & {\tt OBJID, CATNAME, TARGETNAME, TARGETID, RA, DEC, PRIORITY,} \\
& & 7 (\rf) & {\tt MAG\_1 (old \bj), PROGID, MAG\_2 (old \rf), COMMENT}\\
{\tt FSC} & IRAS Faint Source Catalogue sources & 126 & {\tt OBJID, CATNAME, TARGETNAME, TARGETID, RA, DEC, PRIORITY,} \\
& & & {\tt MAG\_1, PROGID, MAG\_2, COMMENT}\\
{\tt RASS} & ROSAT All-Sky Survey candidate AGN & 113 & {\tt OBJID, CATNAME, TARGETNAME, TARGETID, RA, DEC, PRIORITY,} \\
& & & {\tt MAG\_1, PROGID, MAG\_2, COMMENT}\\
{\tt HIPASS} & sources from the HIPASS H{\sc I} survey & 119 & {\tt OBJID, CATNAME, TARGETNAME, TARGETID, RA, DEC, PRIORITY,} \\
& & & {\tt MAG\_1, PROGID, MAG\_2}\\
{\tt DURUKST} & Durham/UKST galaxy survey extension& 78 & {\tt OBJID, CATNAME, TARGETNAME, TARGETID, RA, DEC, PRIORITY,} \\
& & & {\tt MAG\_1, PROGID, MAG\_2}\\
{\tt SHAPLEY} & Shapley supercluster galaxies & 90 & {\tt OBJID, CATNAME, TARGETNAME, TARGETID, RA, DEC, PRIORITY,} \\
& & & {\tt MAG\_1, PROGID, MAG\_2}\\
{\tt DENISI} & DENIS survey galaxies, $I < 14.85$ & 6 & {\tt OBJID, CATNAME, TARGETNAME, TARGETID, RA, DEC, PRIORITY,} \\
 & & & {\tt MAG\_1, PROGID, MAG\_2, COMMENT} \\
{\tt DENISJ} & DENIS survey galaxies, $J< 13.85$ & 5 & {\tt OBJID, CATNAME, TARGETNAME, TARGETID, RA, DEC, PRIORITY,} \\
 & & & {\tt MAG\_1, PROGID, MAG\_2, COMMENT} \\
{\tt AGN2MASS} & 2MASS red AGN survey candidates & 116 & {\tt OBJID, CATNAME, TARGETNAME, TARGETID, RA, DEC, PRIORITY,} \\
 & & & {\tt MAG\_1, PROGID, MAG\_2, MAG\_3} \\
{\tt HES} & Hamburg/ESO survey candidate QSOs & 129 & {\tt OBJID, CATNAME, TARGETNAME, TARGETID, RA, DEC, PRIORITY,} \\
 & & & {\tt MAG\_1, PROGID, MAG\_2} \\
{\tt NVSS} & candidate QSOs from NVSS & 130 & {\tt OBJID, CATNAME, TARGETNAME, TARGETID, RA, DEC, PRIORITY,} \\
 & & & {\tt MAG\_1, PROGID, MAG\_2} \\
{\tt SUMSS} & bright radio sources from SUMSS & 125 & {\tt OBJID, CATNAME, TARGETNAME, TARGETID, RA, DEC, PRIORITY,} \\
 & & & {\tt MAG\_1, PROGID, MAG\_2} \\
\hline \\
\end{tabular}\\
\end{center}
\begin{flushleft}
$\dagger$~~New parameters created for the final data release.\\
{\sl SLANTED FONT} V-spectrum parameters ({\sl \_V}) have matching 
R-spectrum ({\sl \_R}) parameters. \\
\end{flushleft}
\end{table*}

Additional downloads in the form of ASCII files are also available from
the database web site. These include a master catalogue compilation of
all redshifts (from both 6dFGS and the literature), as well as a comma-separated
file of the spectral observations. The latter contains an entry for every
6dFGS observation held by the database (including repeats), regardless of redshift quality.
The master catalogue attempts to assign the best available redshift to those sources
determined to be extragalactic. In the case of repeats, a combined 6dFGS redshift
is obtained by error-weighting ($1/(\Delta\,cz)^2$) those $Q=3,4$ redshifts within 
$5\sigma$ (330\,\kms) of an initial $Q=3,4$ median, thereby excluding blunders. Where 
literature redshifts exist
and are consistent with the 6dFGS redshift, the latter is used in the catalogue. In cases
of disagreement ($> 5\sigma$ difference), the 6dFGS redshift is taken and the
mismatch is flagged. Literature redshifts are use, where they exist, for objects that
6dFGS failed to secure. The master catalogue includes the  {\tt TARGETID} for each
object and the {\tt SPECID} references for each 6dFGS observation contributing to
the final redshift, to facilitate cross-referencing with the 6dFGS database.
Completeness maps (calculated from the revised target lists, after 2MASS and 
SuperCOS magnitude changes) will be made available at a future date.

Table~\ref{tab:parameters} lists a subset of the more commonly-used
database parameters, along with detailed descriptions. New parameters
for this final release are indicated. Users should pay particular
attention to the important differences between parameters which have
similar-sounding names but which are significantly different in purpose.
Examples to note are: (i) {\tt Z}, {\tt Z\_ORIGIN}, {\tt Z\_HELIO}, 
{\tt Z\_INITIAL}, and {\tt Z\_HELIO\_INITIAL}, (ii) {\tt QUALITY} and
{\tt Q\_FINAL}, (iii) ({\tt JTOT}, {\tt HTOT}, {\tt KTOT}), ({\tt J},
{\tt H}, {\tt K}), and ({\tt MAG1}, {\tt MAG2}) (from the {\tt TWOMASS}
table), and, (iv) ({\tt BMAG}, {\tt RMAG}), ({\tt BMAGSEL}, 
{\tt RMAGSEL}) and ({\tt BMAG}, {\tt RMAG}) (from the {\tt SUPERCOS}
table). Table~\ref{tab:parameters} details the differences between them.
\begin{table*}
\begin{center}
\caption{Descriptions of some key parameters in the 6dFGS database.
\label{tab:parameters}
}
\begin{tabular}{lcl}
\hline \\
Parameter & Associated Table(s) & Notes \\
\hline \\
& & \\
{\tt TARGETID} & all & Unique source ID (integer), used to link tables.\\ 
{\tt TARGETNAME} & all & Source name, `g$\#\#\#\#\#\#\#-\#\#\#\#\#\#$'. (Sources observed but not in the original\\
& & target list have the form `c$\#\#\#\#\#\#\#-\#\#\#\#\#\#$'). \\ 
{\tt PROGID} & all & Programme ID (integer), identifying the origin of targets.
${\tt PROGID} \le 8$ for main samples.\\ 
{\tt OBJID} & all except {\tt TARGET} & Unique object ID (integer), assigned to each object in all
input catalogues. \\ 
& and {\tt SPECTRA} & \\ 
& & \\
{\tt BMAG,RMAG} & {\tt TARGET} & New \br\ SuperCOSMOS magnitudes following the revision for 2dFGRS by Peacock, \\
& & Hambly and Read. First introduced for DR2. The most reliable \br\ 6dFGS magnitudes.\\ 
{\tt ZCATVEL,ZCATERR} & {\tt TARGET} & Existing redshifts and errors (\kms) from ZCAT \citep{huchra02} where available.\\
{\tt ZCATREF} & {\tt TARGET} & Code indicating source of ZCAT redshift: `$126x$' for earlier 6dFGS
redshifts (subsequently\\
& & ingested by ZCAT), `$392x$' for ZCAT-ingested 2dFGRS redshifts. The `$x$' in both cases\\
& & holds redshift quality (see {\tt QUALITY} below). {\tt ZCATREF}$\leq99$ for other ZCAT surveys.\\
{\tt BMAGSEL,RMAGSEL} & {\tt TARGET} & Old \br\ SuperCOSMOS magnitudes compiled by W.~Saunders. Never used for selection\\
& & and not intended for science. Previously under {\tt BMAG} and {\tt RMAG} in pre-DR2 releases.\\ 
{\tt TEMPLATECODE}   &   {\tt TARGET}   & Code indicating cross-correlation template: `N' = no redshift, `Z' = ZCAT redshift\\
  &   &  (no template used), `T' = 2dFGRS (no template used), 1 $...$ 9 = 6dFGS template code.\\
& & \\ 
{\tt SPECID} & {\tt SPECTRA} & Unique spectral ID (integer). Different for repeat observations of the same object.\\ 
{\tt Z\_ORIGIN} & {\tt SPECTRA} & Is `C' for most spectra, which come from (c)ombined (spliced) V and R spectral frames. \\
& & Is `V' or `R' for unpaired (orphan) data, as applicable.\\ 
{\tt KBESTR} & {\tt SPECTRA} & Template spectrum ID (integer) used for redshift cross-correlation. \\ 
{\tt Z\_HELIO} & {\tt SPECTRA} & Heliocentric redshift. Corrected by $-40 \kms$ for template
offset if {\tt KBESTR}$=1$ or 7.\\ 
& & The redshift intended for science use. \\
{\tt Z} & {\tt SPECTRA} & Raw measured redshift. Not intended for science use. Also template offset corrected.\\ 
{\tt Z\_INITIAL}\,$\dagger$ & {\tt SPECTRA} & Initial copy of redshift {\tt Z}, uncorrected (e.g. for slit vane shifts). Not for scientific use.\\ 
{\tt Z\_UPDATE\_FLAG}\,$\dagger$ & {\tt SPECTRA} & {\tt Z\_HELIO} corrections: `1' if slit-vane corrected, `2' if template corrected, `3' for both.\\ 
{\tt Z\_HELIO\_INITIAL}\,$\dagger$ & {\tt SPECTRA} & Initial version of {\tt Z\_HELIO}, uncorrected (e.g. for slit vane shifts). Not for science use. \\
{\tt QUALITY} & {\tt SPECTRA} & Redshift quality, $Q$ (integer): `1' for unusable measurements, `2' for possible but unlikely\\
& &  redshifts, `3' for a reliable redshift, `4' for high-quality redshifts, and `6' for \\
& & confirmed Galactic sources. Only {\tt QUALITY}$=3$ or 4 should be used for science. ({\tt QUALITY}\\
& & does {\sl not} measure spectral quality.)\\ 
{\tt Q\_FINAL} & {\tt SPECTRA} & Final redshift quality assigned by software. Not intended for general use. Use {\tt QUALITY}.\\ 
{\tt QUALITY\_INITIAL}\,$\dagger$ &  {\tt SPECTRA}  & Quality value at initial ingest, before database revision. Not for general use.\\
{\tt QUALITY\_UPDATE\_COMM}\,$\dagger$  &  {\tt SPECTRA}   & Explanation of quality value changes during database revision.\\

{\tt TITLE\_V,TITLE\_R} & {\tt SPECTRA} & Observation title from SDS configuration file (consisting of field name and plate number).\\ 
{\tt XTALKFLAG}\,$\dagger$ & {\tt SPECTRA} & Fibre number of a nearby object suspected of spectral cross-talk contamination. `$-1$' if \\
& & object is a contaminator itself. `0' if neither a contaminator nor contaminee.\\
{\tt XTALKSCORE}\,$\dagger$ & {\tt SPECTRA} & Score from `0' (none) to `5' (high) assessing the likelihood of spectral cross-contamination.\\ 
{\tt XTALKVELOFF}\,$\dagger$ & {\tt SPECTRA} & Velocity offset (\kms) between contaminator and contaminee in cross-contamination.\\ 
{\tt XTALKCOMM}\,$\dagger$ & {\tt SPECTRA} & Comment about cross-talk likelihood.\\ 
{\tt SLIT\_VANE\_CORR}\,$\dagger$ & {\tt SPECTRA} & Correction (\kms) made to a redshift affected by slit vane shifts during observing.\\ 
{\tt REVTEMPLATE}\,$\dagger$  &  {\tt SPECTRA}   &   Code of any spectral template used during the database revision of redshifts.\\
{\tt REVCOMMENT}\,$\dagger$ &   {\tt SPECTRA}   &  Explanation of any redshift changes resulting from the database revision.  \\
 & & \\
{\tt CATNAME} & {\tt TWOMASS} & 2MASS name. (Prior to this release, {\tt CATNAME} held the old names now in {\tt PREVCATNAME}).\\
{\tt PREVCATNAME}\,$\dagger$ & {\tt TWOMASS} & Old 2MASS name (as at 2001). \\ 
{\tt RTOT\,}$\dagger$ & {\tt TWOMASS} & 2MASS XSC extrapolated/total radius (2MASS {\tt r\_ext} parameter).\\ 
{\tt JTOT,HTOT,KTOT}\,$\dagger$ & {\tt TWOMASS} & Revised 2MASS XSC total \jhk\ magnitudes 
(2MASS {\tt j\_m\_ext}, {\em etc.}). For science use.\\ 
{\tt MAG\_1,MAG\_2} & {\tt TWOMASS} & Input catalogue magnitudes. Not used in {\tt TWOMASS} table and so default non-value is 99.99.\\ 
& & Superseded by {\tt JTOT}, {\tt HTOT}, and {\tt KTOT}.\\ 
{\tt CORR} & {\tt TWOMASS} & Magnitude correction (based on average surface brightness) used to calculate {\tt KEXT\_K}.\\
{\tt J,H,K} & {\tt TWOMASS} & Old 2MASS XSC total \jhk\ magnitudes. $JH$ used for selection. Superseded by {\tt JTOT}, {\em etc}.\\
{\tt KEXT} & {\tt TWOMASS} & Redundant 2MASS extrapolated $K$ magnitudes, previously used to obtain {\tt KEXT\_K}. \\ 
{\tt KEXT\_K} & {\tt TWOMASS} & Old total $K$ magnitude estimated from {\tt KEXT} and {\tt CORR}. Used in original 6dFGS $K$-band\\
& & selection (see \citet{jones04} for a discussion). Now redundant.\\ 
& & \\
{\tt MAG\_1,MAG\_2} & {\tt SUPERCOS} & Old \br\ SuperCOSMOS magnitudes compiled by Saunders,
Parker and Read for target\\
& & selection. Now superseded by the revised magnitudes {\tt BMAG} and {\tt RMAG} in the {\tt TARGET} table.\\ 
& & \\
\hline \\
\end{tabular}
\end{center}
\begin{flushleft}
$\dagger$~~New parameters created for the final data release.\\
\end{flushleft}
\end{table*}


\subsection{Changes made for the final redshift release}

All of the changes previously implemented for DR2 \citep{jones05} have
been retained, with some modifications. In particular, some fields
rejected from earlier data releases on technical grounds have been fixed
and included in the final release. The final data span observations from
2001 May to 2006 January inclusive. New changes are as follows:
\begin{enumerate}
\item {\bf Revised 2MASS Names:}~~Between the creation of the initial
  6dFGS target list in 2001 and the final 2MASS XSC data release in 2004,
  the 2MASS source designations changed in the last two digits in both
  the R.A.\ and Dec.\ components of the source name. The original 2MASS
  names (previously held in the 6dFGS database {\tt TWOMASS} table under
  the attribute {\tt CATNAME}) have been retained but re-badged under a
  new attribute {\tt PREVCATNAME}. The revised 2MASS names are stored in
  {\tt CATNAME} and are consistent with the final data release of the
  2MASS XSC. Original 6dFGS sources that were subsequently omitted from
  the final 2MASS data release have {\tt CATNAME}\,=\,`\,'.

\item {\bf Revised 2MASS Photometry:}~~The \jhk\ total magnitudes used
  to select 6dFGS sources were also revised by 2MASS between 2001 and
  2004. These new values are held in the newly-created {\tt JTOT}, 
  {\tt HTOT}, {\tt KTOT}. The revisions
  amount to less than 0.03\,mag, except in the case of corrected blunders.
  The old magnitudes used for target selection
  continue to be held in {\tt J}, {\tt H} and {\tt KEXT\_K}, the latter
  being derived from surface brightness-corrected 2MASS
  extrapolated magnitudes (see \citet{jones04} for a full
  discussion).

\item {\bf Revised SuperCOSMOS Photometry:}~~As discussed in
  \cite{jones05} for DR2, the SuperCOSMOS magnitudes were also revised between
  2001 and 2004. As was the case for DR2, {\tt BMAG} and {\tt RMAG} are
  the revised \br\ magnitudes, which should be used for science
  purposes. However, some of the values in {\tt BMAG} and {\tt RMAG}
  have changed because of an improvement in the algorithm we have used
  to match 6dFGS objects with new SuperCOSMOS magnitudes. This has
  removed much more of the deblending discussed in Sec.~2.3 of
  \citet{jones05}. 
  The historical \br\ magnitudes held in {\tt BMAGSEL}, {\tt RMAGSEL}
  (in the {\tt TARGET} table) and {\tt MAG\_1}, {\tt MAG\_2} (in the
  {\tt SUPERCOS} table) retain their DR2 definitions and values.

\item {\bf Redshift Completeness:}~~The 2MASS and SuperCOSMOS magnitude
  revisions have imparted a small but non-negligible scatter between the
  old and new versions of \brjhk, particularly \br$K$. They have a
  non-negligible impact on estimates of 6dFGS redshift completeness at
  the faint end (faintest $\sim0.5$~mag) of each distribution. In this
  regime, the new magnitudes cause increasing numbers of original 6dFGS
  targets to lie beyond the cut-off and increasing numbers of sources
  that were not original targets to fall inside the cut-off.
  Consequently, new target lists were compiled using the revised
  magnitudes, the completeness estimates were recalculated, and the
  results are presented along with the luminosity and mass functions in 
  Jones et~al.\ (in prep.).

\item {\bf Fibre Cross-talk:}~~Instances of fibre cross-talk, in which
  bright spectral features from one spectrum overlap with an adjacent
  one, have been reviewed and are now flagged in the database through
  three new parameters: {\tt XTALKFLAG}, {\tt XTALKSCORE}, and 
  {\tt XTALKVELOFF}, defined in Table~\ref{tab:parameters}. The flags
  are not definitive and are only meant to reflect the {\sl likelihood}
  that a redshift has been affected thus. Specifically, users are urged
  to use extreme caution with redshifts from sources having 
  ${\tt XTALKFLAG} \geq 1$, ${\tt XTALKVELOFF} > 0$ and 
  ${\tt XTALKSCORE} \geq 4$. Cases of ${\tt XTALKSCORE} = 3$ are weak
  candidates where cross-talk is possible but not fully convincing
  (e.g.\ only the V or the R spectra are affected, but not both). 
  ${\tt XTALKSCORE} = 4$ are good candidates, but which carry the
  previous caveat. ${\tt XTALKSCORE} = 5$ are likely cross-talk pairs
  which are usually confirmed through visual inspection of the spectra.
  Cross-talk is an uncommon occurrence (about $\sim 1$ 
  percent of all spectra), and it only affects the redshifts for
  spectra with fewer real features than false ones. An algorithm was
  used to search for coincident emission lines in adjacent spectra and a
  cross-talk severity value assigned from 1 to 5. Users are urged to
  exercise caution with spectra and redshifts having cross-talk values
  of 3 or greater. A detailed discussion of the cross-talk
  phenomenon can be found in the database documentation on the website.

\item {\bf Highest Redshift Sources:}~~Very occasionally, spurious
  features due to cross-talk or poor sky-subtraction led to erroneously
  high redshifts. This is particularly the case with the Additional Target samples
  (${\tt PROGID} > 8$), whose selection criteria do not necessarily
  ensure reliable detections at the optical wavelengths of 6dFGS spectra.
  Special care should be taken with the high redshift sources reported
  for these targets. All sources (across all programmes) with $z \geq 1.0$
  were re-examined and re-classified where necessary. 
  In addition, those sources from the primary and secondary samples
  (${\tt PROGID} \leq 8$) with redshifts in the range $0.2 \leq z < 1.0$
  were re-examined.    
  There are 318 6dFGS sources with $z>1$,
  mostly QSOs, and a further 7 possible cases. The highest of these is
  the $z=3.793$ QSO g2037567$-$243832. Other notable examples are the
  candidate double QSO sources g0114547$-$181903 ($z=2.524$) shown in
  Fig.~\ref{fig:example}($b$) and g2052000$-$500523 ($z=1.036$). Deep
  follow-up imaging in search of a foreground source is necessary to
  decide whether these sources are individual gravitationally lensed
  QSOs or genuine QSO pairs. Even with such data in hand, the
  distinction is quite often equivocal \citep[e.g.][and references
  therein]{faure03,hennawi06}.

\item {\bf Orphan Fields:}~~The final data release includes (for the
  first time) data from 29 orphan fields. These are fields that, for
  various reasons, are missing either the V or R half of the spectrum.
  These fields have a reduced redshift yield because of the restricted
  access to redshifted spectral features, particularly in the case of
  missing R spectra. Orphan field data are flagged in the database
  through the {\tt Z\_ORIGIN} parameter (see
  Table~\ref{tab:parameters}).

\item {\bf Re-examination of Q=1 and Q=2 spectra:}~~All sources originally
classified as either being extragalactic and $Q=2$, or non-2MASS-selected 
(${\tt PROGID} > 4$) and $Q=1$, have been re-examined. This was done primarily
to improve the identification of faint high-redshift QSOs. Many QSOs were poorly 
identified in the early stages of the survey due to the absence of suitable QSO
templates for redshifting. Redshift data for 4\,506 $Q=2$ and 3\,687 $Q=1$ sources were
checked, and the database updated where necessary.

\item {\bf Image Examination of all Q=6 Sources and Re-redshifting:}~~In the initial redshifting
  effort, 6\,212 sources were classified as $Q=6$ (i.e.\ confirmed
  Galactic sources with $z=0$) on the basis of their spectra and
  redshifts alone. Once spectral and imaging data were assembled
  side-by-side in the 6dFGS database, it was straightforward to examine
  the postage-stamp images of these sources, given their spectral
  classification. Most were confirmed as being true Galactic sources
  (stars, H{\sc II} regions, planetary nebulae, YSOs), or Galactic
  objects in close proximity to an extragalactic source. A small number
  were also found to be 2MASS imaging artefacts, or parts of larger
  objects. However, a significant number (847) were found to be
  galaxies with near-zero redshifts, which were
  subsequently re-redshifted and re-classified, and
  updated in the database. 
  In some cases, even though the source was clearly a galaxy on the
  basis of its imaging, its true redshift could not 
  be obtained. The most common causes were scattered light from
  a nearby star, or contamination from a foreground screen of Galactic
  emission.

\item {\bf Anomalous $K$--$z$ Sources with Q=3,4:}~~The $K$--$z$
  magnitude-redshift relation was used to identify anomalous redshifts ($Q=3,4$) outside
  the envelope normally spanned by this relation at typical 6dFGS
  redshifts. The postage-stamp images of these sources were compared to
  their spectra and redshifts to decide if the initial redshift was
  incorrect. There were 120  objects deemed to have an anomalous
  $K$--$z$; 94 were found to have incorrect redshifts,  which 
  were re-examined and re-incorporated into the database.

\item {\bf Correction of Slit-Vane Shifted Fields:}~~Midway through the
  survey it became apparent that the magnetically-held vane supporting
  the spectrograph slit was shifting occasionally between exposures.
  This problem was discovered prior to DR2 but the affected redshifts
  were withheld; they have been corrected and provided in the final
  release. The resulting spectra from affected fields show a small
  wavelength offset (greater than $\pm0.75$\AA\ and up to a few \AA),
  dependent on fibre number. The V and R spectral halves were sometimes
  affected individually, and at other times in unison. Instances of
  shifting were isolated by comparing the wavelength of the 
  [{\sc OI}]$\lambda$5577.4\AA\ sky line, as measured from the 6dFGS
  spectra, to its true value. A search found 125 affected fields able to
  be satisfactorily fit (measured [{\sc OI}] against fibre number) and
  redshift corrected. In all, 18\,438 galaxies were corrected in
  this way (approximately 14 percent of the entire sample of 
  {\sl all} spectra), with corrections $\lesssim 12$\AA. Redshift
  template values {\tt KBESTR} were used to determine whether to apply a
  correction. If an object used ${\tt KBESTR}=1,2$ (corresponding to
  early-type galaxy templates), the redshift was deemed to be due to
  absorption-lines, which occur predominantly in the V half. If the
  corresponding V frame was indeed slit-vane affected, a correction was
  applied to the redshift for this galaxy based on the fit to the V
  frame {\em alone}. Alternatively, if ${\tt KBESTR}=3,4,5$
  (corresponding to late-type galaxy templates), then the redshift was
  deemed to be emission-line dependent, and the corresponding R frame
  correction was made where necessary. Users can find those galaxies
  with slit-vane corrected redshifts through the new {\tt SLITVANECORR}
  parameter, which gives the size (in \kms) of any corrections applied.
  Unaffected galaxies have ${\tt SLITVANECORR} = 0$. The corrected
  redshifts are the heliocentric redshifts held by {\tt Z\_HELIO}.

\item {\bf Correction for Template Offset Values:}~~Various tests
  comparing 6dFGS redshifts to independent measurements found small
  systematic offsets in the case of a couple of templates. The
  discrepancy is almost certainly due to a zero-point error in the
  velocity calibration of the template spectra. This effect was
  discovered prior to DR2 and is discussed in \citet{jones05}, although
  no corrections were applied to the affected redshifts in that release.
  For this final release, corrections of $-40$\,\kms\ have been applied
  to redshifts derived from templates ${\tt KBESTR}=1,7$. The corrected
  redshifts are both the raw ({\tt Z}) and heliocentric ({\tt Z\_HELIO})
  redshifts. The redshift offsets were found to be consistent between a
  2004 comparison of 16\,127 6dFGS and ZCAT redshifts, and a 2007
  comparison of 443 redshifts from various peculiar velocity surveys
  \citep{bernardi03, smith00, smith04, wegner03}.

\item {\bf Telluric Sky Line Subtraction:}~~The redshifting software
  used by 6dFGS automatically removed telluric absorption lines from
  spectra, but the database spectra have hitherto retained their
  imprint. For the final release we have re-spliced spectra and
  incorporated telluric line removal. An example spectrum is shown in
  Fig.~\ref{fig:example}. A small number of spectra which failed to
  re-splice successfully have had their old telluric-affected versions
  retained.
  
\item {\bf Spurious Clustering:}~~The entire sample of reliable redshifts
($Q=3,4$) was tested for spurious clusters, caused by any systematic
effect that produces noticeable numbers of objects from the same field
with nearly identical redshifts. Possible causes include poor sky subtraction
and/or splicing of spectra, and the fibre cross-talk effect discussed in item
(v). Fields containing at least 16 cases of galaxy groups (3 or more members)
 with redshift differences of less than 30\,\kms\ had their redshifts re-examined:
 171 galaxies from 7 fields.  No prior knowledge of 
 real galaxy clustering was used for the re-redshifting, and the database was 
 updated with new redshifts and quality assignments.
 The field 0058m30 was particularly prominent with
 48 galaxies at or near an apparent redshift of 0.1590. This was due to the
 over-subtraction and subsequent mis-identification of the 7600\,\AA\  telluric 
 absorption band with redshifted H$\alpha$. A further 134 objects with 
 redshifts in the range $0.1585 \leq z \leq 0.1595$ were reexamined for this
 effect, and 118 given corrected $z$ or $Q$ values. Almost all of the affected
 spectra are among the earliest observations of survey data (2001), prior 
 to the switch to VPH gratings.

\item {\bf RASS sources:}~~All sources in the ROSAT All-Sky Survey (RASS)
Additional Target sample (${\tt PROGID} = 113$; 1850 sources) were 
re-examined using the full QSO template set. The database was updated with
new redshifts and quality assignments. \citet{mahoney09} describe the
selection and characteristics of this sample in more detail.

\end{enumerate}

\begin{figure*}

\plotfull{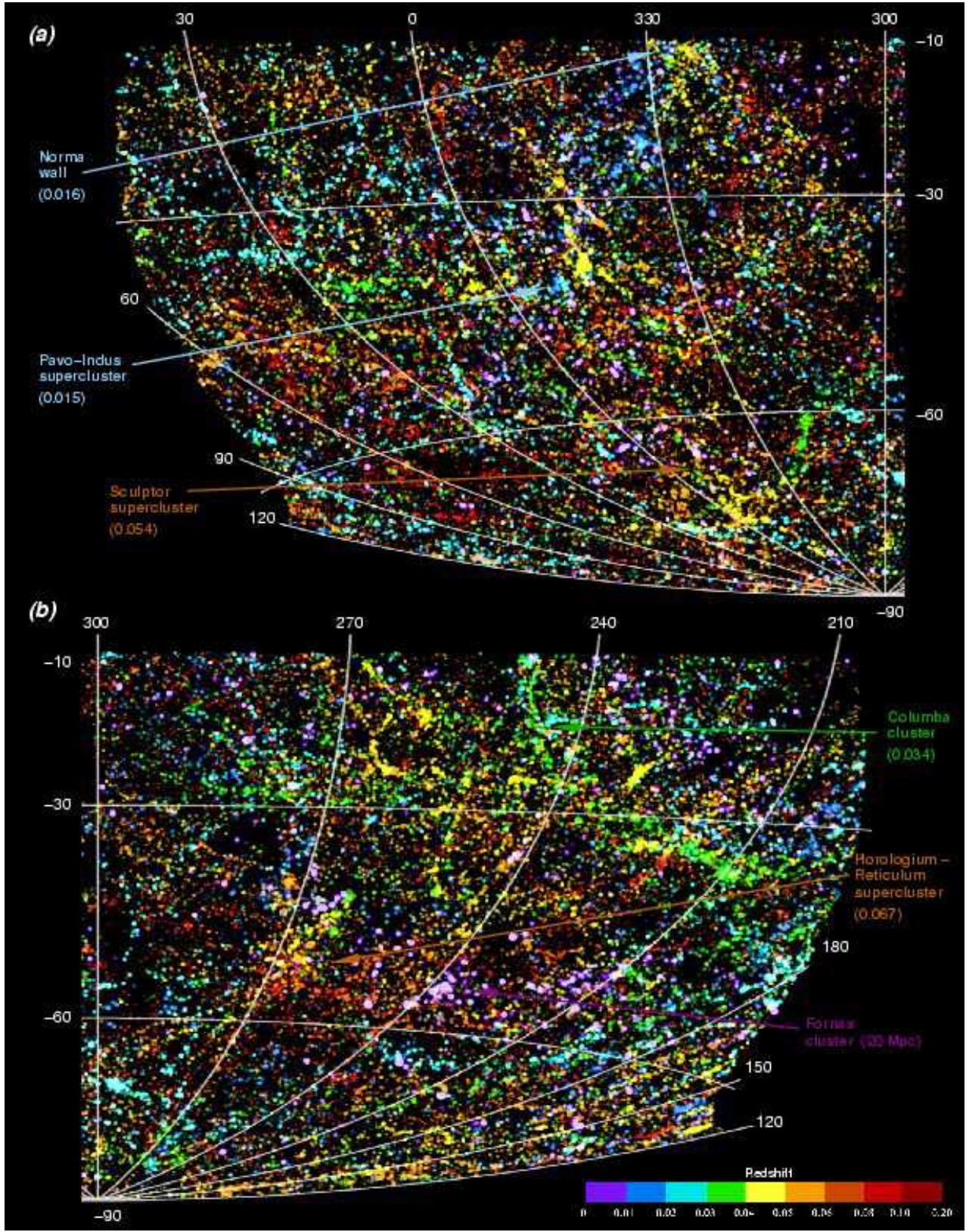}{1.0}

\caption{The distribution of galaxies in the 6dFGS shown in an Aitoff
projection of Galactic coordinates across the southern Galactic
hemisphere; redshifts are colour-coded from blue (low, $z < 0.02$) to red (high, $z>0.1$).
Some of the major large-scale structures are labelled.}
\label{fig:colplot1}
\end{figure*}

\begin{figure*}

\plotfull{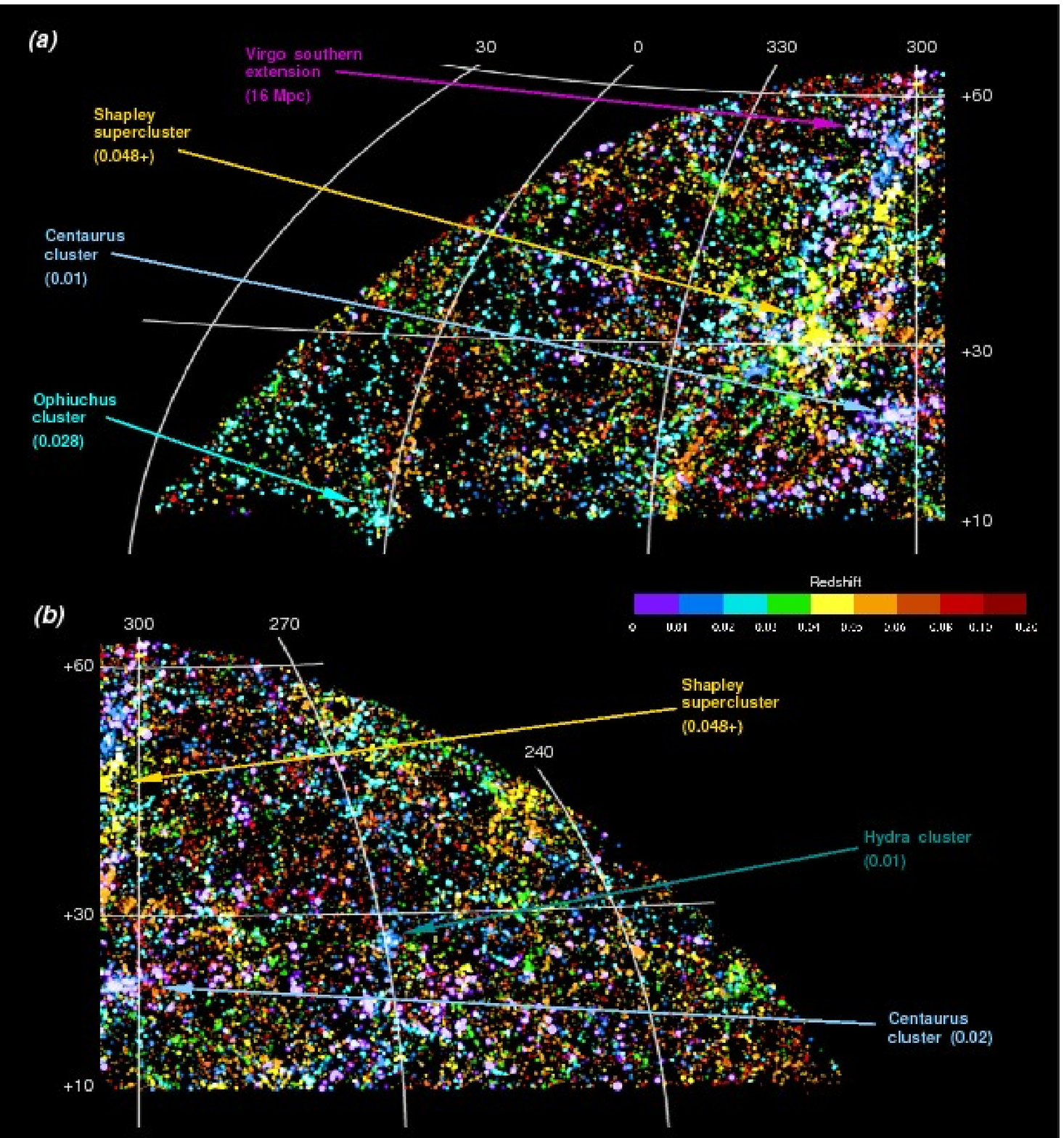}{1.0}

\caption{Same as Fig.~\ref{fig:colplot1} except showing the northern
  Galactic hemisphere.}
\label{fig:colplot2}
\end{figure*}


\section{Southern Large-Scale Structures}
\label{sec:results}

\subsection{Sky projections}

The wide sky coverage of the 6dF Galaxy Survey affords the most detailed
view yet of southern large-scale structures out to $cz\sim$30\,000~\kms.
The 6dFGS extends the sky coverage of the 2dFGRS \citep{colless01}
by an order of magnitude, and likewise improves by an order of magnitude
on the sampling density of the all-sky PSCz survey \citep{branchini99,saunders00}.
Prominent southern structures such as Shapley, Hydra-Centaurus and
Horologium-Reticulum have received much special attention in their own
right over recent years \citep{raychaudhury89, quintana95,
drinkwater99, reisenegger00, bardelli00, kaldare03, woudt04,
fleenor05, fleenor06, radburnsmith06, proust06}. However, a detailed
large-scale mapping of all intervening structures (and the voids between
them) with a purpose-built instrument has remained unavailable until
now. The complementary 2MASS Redshift Survey (2MRS; Huchra et al., in
prep) uses the 6dFGS in the south to provide an all-sky redshift survey
of some 23\,000 galaxies to $K=11.25$ ($\bar{z}=0.02$). It is hoped it
will one day be extended to reach an equivalent depth to 6dFGS in the
north in those areas not already covered by SDSS.

Figures~\ref{fig:colplot1} and~\ref{fig:colplot2} show the $z<0.2$
universe as seen by 6dFGS in the plane of the sky, projected in Galactic
coordinates. The two figures show the northern and southern Galactic
hemispheres, respectively. Familiar large-scale concentrations such as
Shapley are obvious, and several of the key structures have been labelled. 
At $z<0.02$, filamentary structures such as the Centaurus, Fornax and
Sculptor walls \citep{fairall98} interconnect
their namesake clusters in a manner typical of large structures
generally. At $z\approx0.006$ to 0.01 the Centaurus wall crosses the
Galactic plane Zone of Avoidance (ZoA) and meets the Hydra wall at the
Centaurus cluster. The Hydra wall then extends roughly parallel to the
ZoA before separating into two distinct filaments at the adjacent
Hydra/Antlia clusters, both of which extend into the ZoA. Behind these,
at $z=0.01$ to 0.02, a separate filament incorporates the Norma and
Centaurus-Crux clusters, and encompasses the putative Great Attractor
region \citep[][and references therein]{woudt04,radburnsmith06}. Beyond
these, at $z=0.04$ to 0.05, lies the Shapley Supercluster complex, a
massive concentration of clusters thought to be responsible for 10
percent of the Local Group motion \citep{raychaudhury89, reisenegger00,
bardelli00} or even more \citep{quintana95, drinkwater99, proust06}.


\subsection{Declination slice projections}

\begin{figure*}

\plotfull{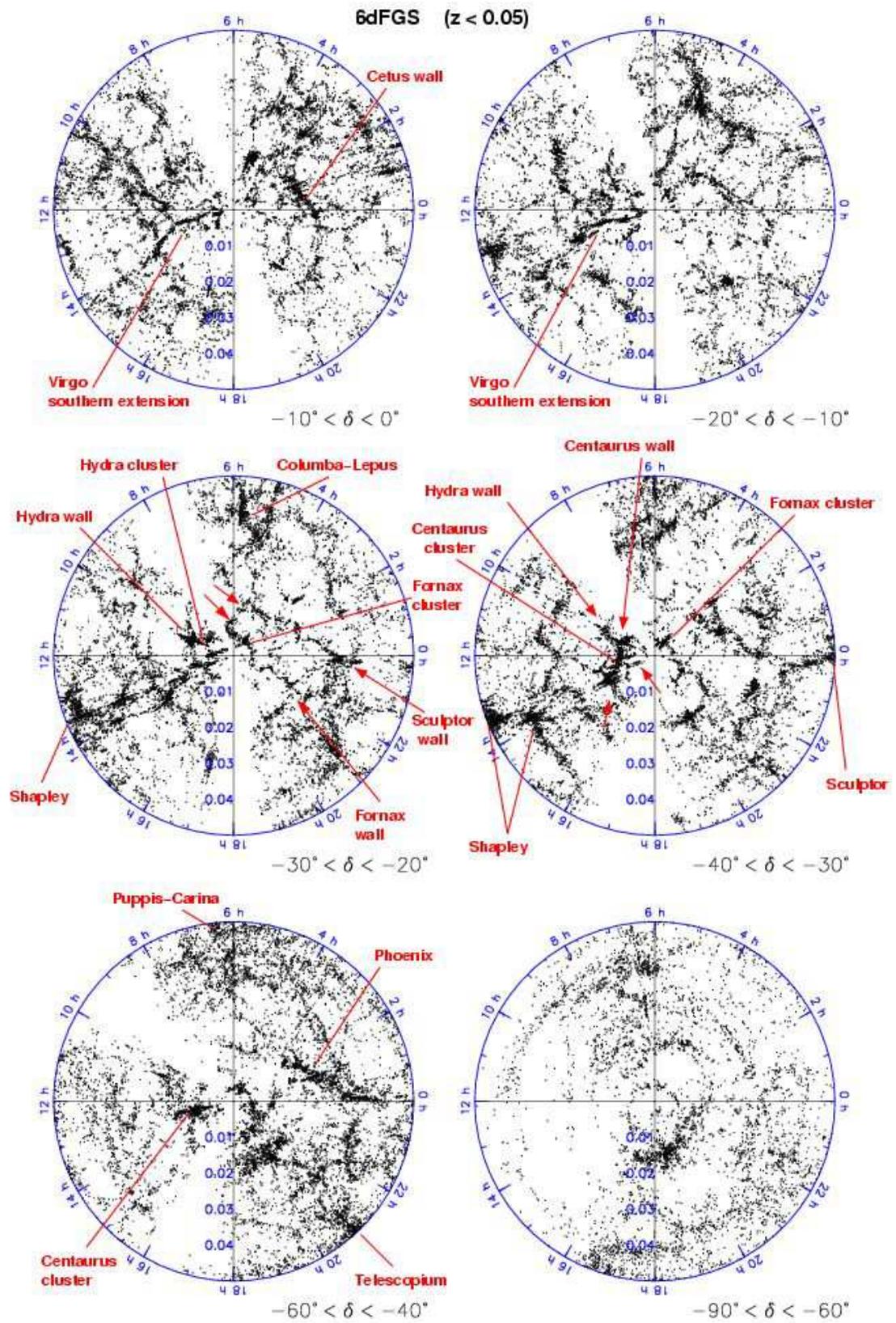}{0.88}

\caption{6dFGS redshift maps out to $z=0.05$, in declination slices of varying width
from the equator to the pole.}
\label{fig:pieinner}
\end{figure*}

\begin{figure*}

\plotfull{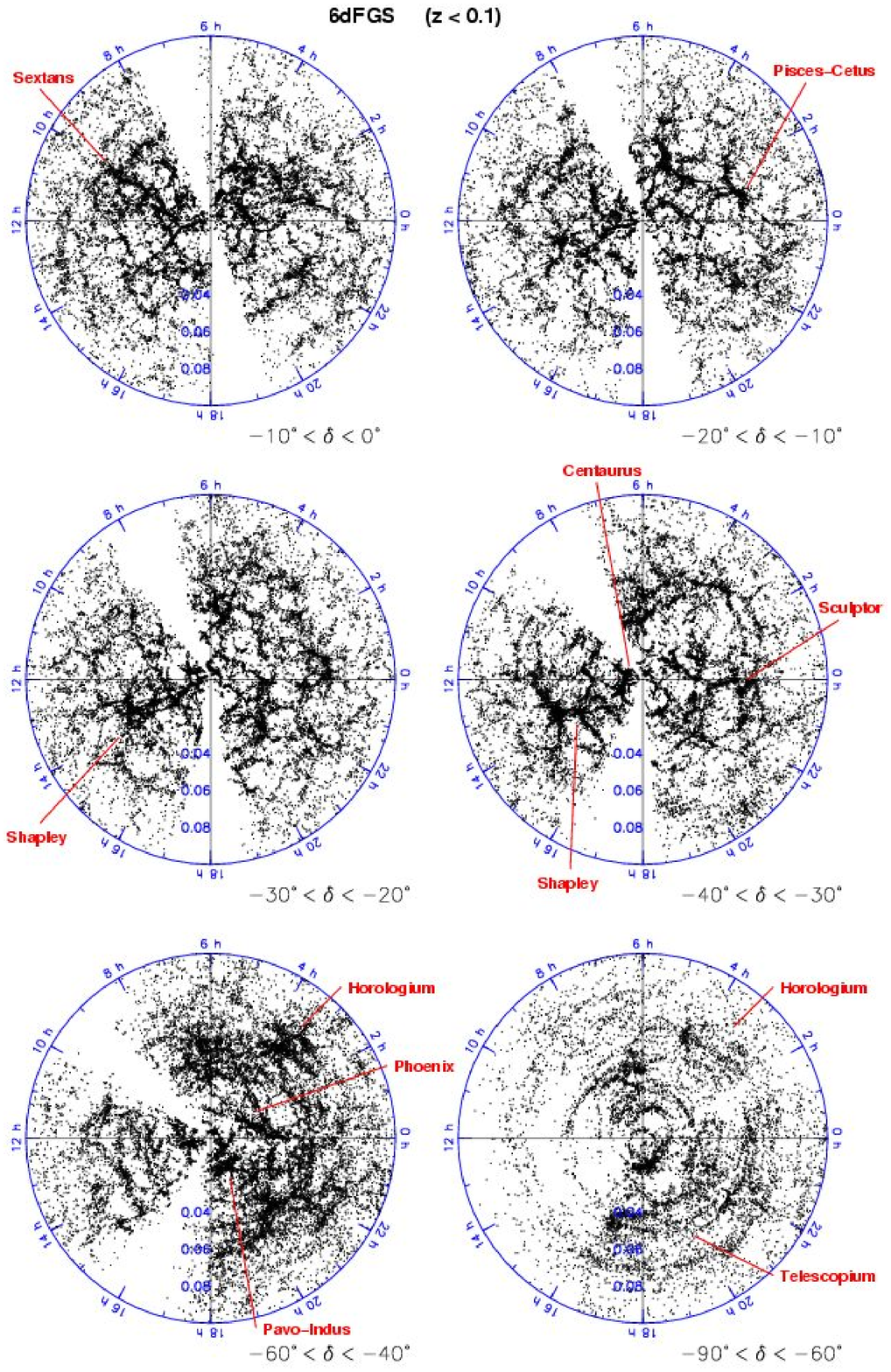}{0.88}

\caption{Same as Fig.~\ref{fig:pieinner}, except on a larger scale out
  to $z=0.1$.}
\label{fig:pieouter}
\end{figure*}

Figures~\ref{fig:pieinner} and~\ref{fig:pieouter} show an alternative
projection of these structures, as conventional radial redshift maps,
cross-sectioned in declination. The two figures show the same data on
two different scales, out to limiting redshifts of $z=0.05$ and 0.1
respectively. The empty sectors in our maps correspond to the ZoA
region. These declination-slice sky
views can also be cross-referenced with the Aitoff-projected sky
redshift maps presented in \citet{jones05} for the 6dFGS data available
up to 2004, as well as Figs~\ref{fig:colplot1} and~\ref{fig:colplot2}.

Figure~\ref{fig:pieouter} similarly displays the local universe out to
$z = 0.1$ with hitherto unseen detail and sky coverage. While it extends
and confirms the now familiar labyrinth of filaments and voids, it also reveals 
evidence of inhomogeneity on a still larger scale --- the plot for $-40^\circ < \delta
< -30^\circ$ (middle right panel) is a good example. A large under-dense
region ($\Delta z \sim 0.05$) at $\alpha \approx 4$\,hr to 5\,hr separates
regions of compact high-density filaments; similar inhomogeneities are
visible in the other plots. An extraordinarily large void ($\Delta z = 0.03$
by 0.07) is apparent in the plot for $-20^\circ < \delta <
-10^\circ$, towards $\alpha \approx 23$\,hr. Other voids of this size
are apparent when the data are examined in Cartesian coordinates. The
most extreme inhomogeneity, however, is the over-dense Shapley region,
which is unique within the sample volume.

\citet{erdogdu06a} have used spherical harmonics and Wiener filtering to
decompose the density and velocity field of the shallower 2MRS. The correspondence
between the largest-scale superclusters and voids seen in both surveys
at $z<0.05$ is clear. Our southernmost projection ($-90^\circ < \delta <
-60^\circ$) confirms the most distant (Pavo) of the three tentative superclusters
of \citet{fairall06} while indicating that the other two are not major overdensities.
We point out that this southern region is where 6dFGS coverage is generally
lowest, with below-average completeness between 0~hr and 6~hr and around the pole
(poor sky coverage), and at 11~hr to 17~hr (ZoA). Azimuthal stretching effects
are also evident, due to the wide R.A.\ span of single fields at
polar declinations.

Work is currently underway cataloguing new clusters and groups from
6dFGS (Merson et~al, in prep.) using a percolation-inferred friends-of-friends algorithm
\citep{huchra82,eke04}.
At the same time, a preliminary list of $\sim500$ void regions has been
compiled as a reference for future work on under-dense regions. A power
spectrum analysis of the clustering of 6dFGS galaxies will
be published elsewhere.


\section{Conclusion}
\label{sec:conclusion}

The 6dF Galaxy Survey (6dFGS) is a combined redshift and peculiar velocity
survey over most of the southern sky. Here we present the final redshift
catalogue for the survey (version 1.0), consisting of 
125\,071 extragalactic redshifts over the whole southern sky with
$|b|>10^\circ$. Of these, 110\,256 are new redshifts from 136\,304 spectra
obtained with the United Kingdom Schmidt Telescope (UKST) between
2001 May and 2006 January. With a median redshift of $z_{1/2}=0.053$, 
6dFGS is the deepest hemispheric redshift survey to date. 
Redshifts and associated spectra are available through a
fully-searchable online SQL database, interlinked with photometric and
imaging data from the 2MASS XSC, SuperCOSMOS and a dozen other input
catalogues. Peculiar velocities and distances for the brightest
10~percent of the sample will be made available in a separate future
release.

In this paper we have mapped the large-scale structures of the local
($z<0.1$) southern universe in unprecedented detail. In addition to
encompassing well-known superclusters such as Shapley and
Hydra-Centaurus, the 6dFGS data reveal a wealth of new intervening
structures. The greater depth and sampling density of 6dFGS compared to
earlier surveys of equivalent sky coverage has confirmed hundreds of
voids and furnished first redshifts for around 400
southern Abell clusters \citep{abell89}. More detailed quantitative analyses 
of 6dFGS large-scale structure will be the subject of future publications.

The unprecedented combination of angular coverage and depth in 6dFGS
offers the best chance yet to minimise systematics in the determination
of the luminosity and stellar mass functions of low-redshift galaxies,
both in the near-infrared and optical \citep[e.g.][]{jones06}. While surveys containing 
$\sim10^5$-galaxy redshifts (such as 6dFGS) have now
reduced random errors to comparable levels of high precision, systematic
errors remain the dominant source of the differences between surveys.
For example, the evolutionary corrections that initially beset
comparisons between 2dFGRS and SDSS \citep[cf.][]{norberg02,blanton01}
are negligible for 6dFGS, which spans lookback times of only 0.2 to
0.7~Gyr across $[0.5\bar{z}, 1.5\bar{z}]$ (compared to 0.5 to 1.3~Gyr
for SDSS and 2dFGRS). The minimisation of such systematics are a feature
of the 6dFGS stellar mass and luminosity functions derived for the final
redshift set (Jones et~al., in prep).

In addition to these studies, 6dFGS redshift data have already been used
to support a variety of extragalactic samples selected from across the
electromagnetic spectrum. Deep H{\sc I} surveys planned for
next-generation radio telescopes \citep{blake04,vandriel05,rawlings06,johnston08} 
will also benefit from this redshift information as they probe the gas content of the local
southern universe over comparable volumes.


\section*{Acknowledgements}

DHJ acknowledges support from Australian Research Council
Discovery--Projects Grant (DP-0208876), administered by the Australian
National University. JPH acknowledges support from the US National Science
Foundation under grant AST0406906.

We dedicate this paper to two colleagues who made important contributions to the 
6dF Galaxy Survey before their passing: John Dawe (1942 -- 2004),
observer and long-time proponent of wide-field fibre spectroscopy on the UKST from
its earliest days, and Tony Fairall (1943 -- 2008), whose unique insights from a 
career-long dedication to mapping the southern universe underpin much of the 
interpretation contained herein.


\end{document}